\documentclass[]{aa}

\usepackage{graphicx}
\usepackage{amsmath}
\usepackage{txfonts}
\usepackage{textcomp, gensymb}
\usepackage{natbib}
\bibpunct{(}{)}{;}{a}{}{,} 
\usepackage{multirow}
\begin{document}

   \title{High-temperature measurements of acetylene VUV absorption cross sections and application to warm exoplanet atmospheres}


   \author{Benjamin Fleury\inst{1}
          \and Mathilde Poveda\inst{2,3}
          \and Yves Benilan\inst{2}
          \and Roméo Veillet\inst{1}
          \and Olivia Venot\inst{1}
          \and Pascal Tremblin\inst{3}
          \and Nicolas Fray\inst{2}
          \and Marie-Claire Gazeau\inst{2}
          \and Martin Schwell\inst{2}
          \and Antoine Jolly\inst{2}
          \and Nelson de Oliveira\inst{4}
          \and Et-touhami Es-sebbar\inst{5}
          }

   \institute{Université Paris Cité and Univ Paris Est Creteil, CNRS, LISA, F-75013 Paris, France\\
              \email{benjamin.fleury@lisa.ipsl.fr}
         \and
             Univ Paris Est Creteil and Université Paris Cité, CNRS, LISA, F-94010 Créteil, France\\
         \and 
            Université Paris-Saclay, UVSQ, CNRS, CEA, Maison de la Simulation, 91191, Gif-sur-Yvette, France\\
        \and 
            Synchrotron SOLEIL, L’Orme des Merisiers, 91192 Gif sur Yvette, France\\
        \and
            Clean Energy Research Platform (CERP), Physical Sciences and Engineering Division, King Abdullah University of Science and Technology (KAUST), Thuwal 23955-6900, Saudi Arabia\\}

   \date{Received July 24, 2024; accepted November 21, 2024}

 
  \abstract
   {Most observed exoplanets have high equilibrium temperatures (T$_{\mathrm{eq}}$ > 500~K). Understanding the chemistry of their atmospheres and interpreting their observations requires the use of chemical kinetic models including photochemistry. The thermal dependence of the vacuum ultraviolet (VUV) absorption cross sections of molecules used in these models is poorly known at high temperatures, leading to uncertainties in the resulting abundance profiles.}
   {The aim of our work is to study experimentally the thermal dependence of VUV absorption cross sections of molecules of interest for exoplanet atmospheres and provide accurate data for use in atmospheric models. This study focuses on acetylene (C$_{2}$H$_{2}$).}
   {We measured absorption cross sections of C$_{2}$H$_{2}$ at seven temperatures ranging from 296 to 773~K recorded in the 115-230~nm spectral domain using VUV spectroscopy and synchrotron radiation. These data were used in our 1D thermo-photochemical model, to assess their impact on the predicted composition of a generic hot Jupiter-like exoplanet atmosphere.}
   {The absolute absorption cross sections of C$_{2}$H$_{2}$ increase with temperature. This increase is relatively constant from 115 to 185~nm and rises sharply from 185 to 230~nm. The abundance profile of C$_{2}$H$_{2}$ calculated using the model shows a slight variation, with a maximum decrease of 40\% near 5 $\times$ 10$^{-5}$ bar, when using C$_{2}$H$_{2}$ absorption cross sections measured at 773~K compared to those at 296~K. This is explained by the absorption, higher in the atmosphere, of the actinic flux from 150 to 230~nm due to the increase in the C$_{2}$H$_{2}$ absorption in this spectral range. This change also impacts the abundance profiles of other by-products such as methane (CH$_{4}$) and ethylene (C$_{2}$H$_{4}$).}
   {We present the first experimental measurements of the VUV absorption cross sections of C$_{2}$H$_{2}$ at high temperatures. Similar studies of other major species are needed to improve our understanding of exoplanet atmospheres.}


   \titlerunning{High-temperature measurements of acetylene VUV absorption cross section}
   \authorrunning{B. Fleury et al.}
   \maketitle
%

\section{Introduction} \label{sec:Introduction}

With current observational constraints, a large fraction of observed exoplanets whose atmospheres can be studied with spectroscopy orbits close by their host star, receiving a high stellar flux and exhibiting elevated equilibrium temperatures (T$_{\mathrm{eq}}$ > 500~K). In particular, the high flux received in the ultraviolet (UV) has important implications for the composition of the atmosphere of these planets. Indeed, vacuum ultraviolet (VUV) photons (i.e., photons with wavelengths shorter than 200~nm) modify the composition of planetary atmospheres by photodissociating atmospheric constituents, and thus initiating photochemical reactions \citep{Moses.2014}. While this process has been expected and predicted by photochemical models \citep{Lineetal.2010, Lineteal.2011, Mosesetal.2011, Venotetal.2015, Venotetal.2020, Baeyensetal.2022}, only recent observations of the warm gas giants WASP-39b and WASP-107b by JWST have provided the first direct evidence of the presence of photochemistry and its impact on the composition of observed exoplanet atmospheres \citep{Tsaietal.2023, Dyreketal.2024}. Indeed, spectral signatures assigned to sulfur dioxide (SO$_2$) \citep{Aldersonetal.2023, Dyreketal.2024, Powelletal.2024} have been observed for the first time in these atmospheres with JWST. The presence of this compound has been explained by 1D thermo-photochemical models, as a result of a consecutive series of chemical reactions initiated by photochemistry \citep{Tsaietal.2023, Dyreketal.2024}. Therefore, it is now obvious that understanding exoplanet atmospheres and interpreting their observations requires the use of kinetic models that incorporate photochemistry.

To implement photochemistry, 1D thermo-photochemical models require the VUV absorption cross sections of molecules present in exoplanetary atmospheres as input parameters. These data are essential for calculating radiative transfer through the atmosphere and subsequently determining molecular photodissociation rates as a function of altitude. To obtain accurate results, these models need to use data suitable for the appropriate conditions (i.e., temperature, pressure, etc.) encountered in these environments, which are often not comparable to the Solar System. Indeed, uncertainties in these data lead to uncertainties in the molecular abundances predicted by atmospheric models \citep{Venotetal.2013, venotetal2018, Ranjanetal.2020}. In general, the thermal dependence of the VUV absorption cross sections of molecules is poorly known for the range of temperature observed in exoplanet atmospheres. Indeed, most of the corresponding studies are limited to the relatively low temperatures (T < 400~K) of the atmospheres of the Solar System but do not extend to the higher temperatures observed in extrasolar atmospheres (500 – 2000~K). These studies concern various molecules such as SO$_{2}$ \citep{Wuetal.2000, Rufusetal.2009}, methane (CH$_{4}$) and ethane (C$_{2}$H$_{6}$) \citep{ChenandWu.2004}, ethylene (C$_{2}$H$_{4}$) \citep{Wuetal.2004}, or ammonia (NH$_{3}$) \citep{Chengetal.2006, Wuetal.2007}. In addition, studies have also been conducted to measure the absorption cross sections of some molecules such as acetylene (C$_{2}$H$_{2}$) \citep{Vattulainenetal.1997, Zabetietal.2017}, ammonia (NH$_{3}$) \citep{Wengetal.2021}, hydrogen sulfide (H$_{2}$S), carbon disulfide (CS$_{2}$), and carbonyl sulfide (OCS) \citep{Groschetal.2015} at elevated temperatures up to 1500~K, but these measurements are limited to wavelengths longer than 200~nm. However, \citet{Venotetal.2013, venotetal2018} have measured the absorption cross section of carbon dioxide (CO$_2$) across a wider range of temperatures (150 to 800~K) and in a wavelength range critical for photochemistry (115-230~nm). Their studies show that its absorption cross section varies by several orders of magnitude with the temperature. Overall, this demonstrates the need to measure the absorption cross sections of molecules of interest for exoplanetary atmospheres at higher temperatures, as has been highlighted in collaborative white papers \citep{Fortneyetal.2019, Chubbetal.2024}.

In this study, we investigate the thermal dependence of the VUV absorption cross section of acetylene. Although C$_{2}$H$_{2}$ detection has been claimed in only one observation of the atmosphere of HD 209458b \citep{Giacobbetal.2021}, theoretical models predict that it is one of the major products of methane photochemistry in gas giant exoplanet atmospheres \citep{Mosesetal.2013b, Venotetal.2015, KawashimaandMasahiro.2018, Baeyensetal.2022}, making this molecule a possible indicator of disequilibrium chemistry. Moreover, the C$_{2}$H$_{2}$ mixing ratio could be used as an indicator of the atmospheric carbon-to-oxygen ratio (C/O), as the abundance of this molecule is expected to increase significantly in atmospheres with C/O > 1 \citep{Mosesetal.2013a, Venotetal.2015, Rocchettoetal.2016, Drummondetal.2019}.

Numerous experimental measurements of the UV absorption cross section of C$_{2}$H$_{2}$ have been conducted, covering a wide range of wavelengths in the VUV (from 110 to 200~nm) as well as in the mid-UV and near-UV up to 400~nm \citep{Nakayama.1964, FooandInes.1973, Watsonetal.1982, Sutoetal.1984, VanCraenetal.1985, VanCraenetal.1986, Wuetal.1989, Chenetal.1991, Smithetal.1991, Vattulainenetal.1997, Benilanetal.2000, Wuetal.2001, Boyeetal.2004, Chengetal.2011, Zabetietal.2017}. Table~\ref{Table1} provides a summary of the temperature, spectral range, and resolution used in some of these previous studies. 

\begin{table*}[h]
    \centering
        \caption{VUV and UV absorption cross sections of C$_{2}$H$_{2}$ previously reported in the literature.}
    \begin{tabular}{cccc}
    \hline\hline
      Temperature (K) & Spectral range (nm) & Resolution (nm) & References \\ 
      \hline
      85 & 110 - 155	& 0.02 & \citet{Chengetal.2011} \\
      150 & 120 - 215	& 0.007 & \citet{Wuetal.2001} \\
      \multirow{2}{*}{155} & 153 - 193 & 0.08 & \citet{Wuetal.1989} \\
       & 140 - 210	& 0.007 & \citet{Chenetal.1991} \\
       173 & 185 - 235	& 0.02 & \citet{Benilanetal.2000} \\
       193 & 190 - 247 & 0.0001 & \citet{Watsonetal.1982} \\
       195 & 147 - 201 & < 0.0075 & \citet{Smithetal.1991} \\
       \multirow{13}{*}{Ambient} & 106 - 180 & 0.1 & \citet{Nakayama.1964} \\
       & 165 - 195	&  & \citet{FooandInes.1973} \\
       & 190 - 247	& 0.0001 & \citet{Watsonetal.1982} \\
       & 105 - 155	& 0.04 & \citet{Sutoetal.1984} \\
       & 193 - 219	& 0.0001 & \citet{VanCraenetal.1985, VanCraenetal.1986} \\
       & 153 - 193	& 0.08 & \citet{Wuetal.1989} \\
       & 137 - 201 & < 0.0075 & \citet{Smithetal.1991} \\
       & 200 - 400	& 0.5 & \citet{Vattulainenetal.1997} \\
       & 185 - 235	& 0.02 & \citet{Benilanetal.2000} \\
       & 120 - 230 & 0.007 & \citet{Wuetal.2001} \\
       & 105 - 153	& 0.08 & \multirow{2}{*}{\citet{Boyeetal.2004}} \\
       & 116.9 - 117.5	& 0.0012 & \\
       & 110 - 155	& 0.02 & \citet{Chengetal.2011} \\
       & 200 - 300	& 3.5 & \citet{Zabetietal.2017} \\
    370 & 120 - 140 & 0.007 & \citet{Wuetal.2001} \\
    873, 1073 & 200 - 400 & 0.5 & \citet{Vattulainenetal.1997} \\
    565 - 1500 & 200 - 300 & 3.5 & \citet{Zabetietal.2017} \\
      \hline
    \end{tabular}
    \label{Table1}
\end{table*}

As is shown in Table~\ref{Table1}, most of these measurements have been conducted at ambient temperature ($\sim$ 295~K) or at lower temperatures relevant to the atmospheres of other Solar System objects where C$_{2}$H$_{2}$ has been observed, such as the gas giants \citep{ridgwayetal.1974, Moos1979, Orton1987, Herbertetal1987, Macy1980}, Titan \citep{Horst2017, Nixon2024} and Pluto \citep{Sternetal2015}. Only a few measurements have been conducted at higher temperatures. \citet{Wuetal.2001} measured the absorption cross section of C$_{2}$H$_{2}$ from 120 to 140~nm at 370~K and observed an average increase of 20\% in the absorption cross section compared to data at 295~K. \citet{Vattulainenetal.1997} measured the absorption cross section of C$_{2}$H$_{2}$ between 200 and 400~nm at ambient temperature, 873~K, and 1073~K and showed an increase of up to a factor of ten in the cross section with temperature. Finally, \citet{Zabetietal.2017} have measured the absorption cross section from 200 to 300~nm, of shocked heated C$_{2}$H$_{2}$ in the 565 – 1500~K temperature range. The authors also observed an increase in the absorption cross section with temperature as well as a shift of the maximum of the absorption toward longer wavelengths. There is, therefore, no existing data on the absorption cross section of C$_{2}$H$_{2}$ for both the range of temperatures and the whole VUV spectral domain that would be relevant for modeling the photochemistry in exoplanet atmospheres.

In this paper, we present a study of the thermal dependence of the VUV absorption cross section of C$_{2}$H$_{2}$ from 296 to 793~K and for a large spectral domain ranging from 115 to 230~nm. Thus, our study encompasses most of the spectral domain of C$_{2}$H$_{2}$ photodissociation, which extends from the photodissociation threshold at 5.71~eV (217~nm) to the photoionization threshold at 11.27~eV (110~nm) \citep{Heaysetal.2017}. In Sect. \ref{sec:Exp. Methods and Protocols}, we describe the new UV spectroscopy platform used to determine the absorption cross section of C$_{2}$H$_{2}$. Our measurements, their analyses, and the atmospheric simulations to quantify the impact of these new data on the abundances predicted by thermo-photochemical models are presented in Sect. \ref{sec:Results and discussions}. Finally, the conclusions of this study are summarized in Sect. \ref{sec:Conclusions}. The C$_{2}$H$_{2}$ absorption cross section data measured in this study can be downloaded from the ExoMol database \citep{Tennysonetal2024}\footnote{https://www.exomol.com} and the ANR EXACT project website\footnote{https://www.anr-exact.cnrs.fr/fr/absorption-cross-sections}.

\section{Experimental methods and protocols} \label{sec:Exp. Methods and Protocols}
\subsection{Measurements of acetylene UV absorption spectra} \label{subsec:Measurements of acetylene UV absorption spectra}

\begin{figure*}[h]
    \centering
    \includegraphics [width=\hsize] {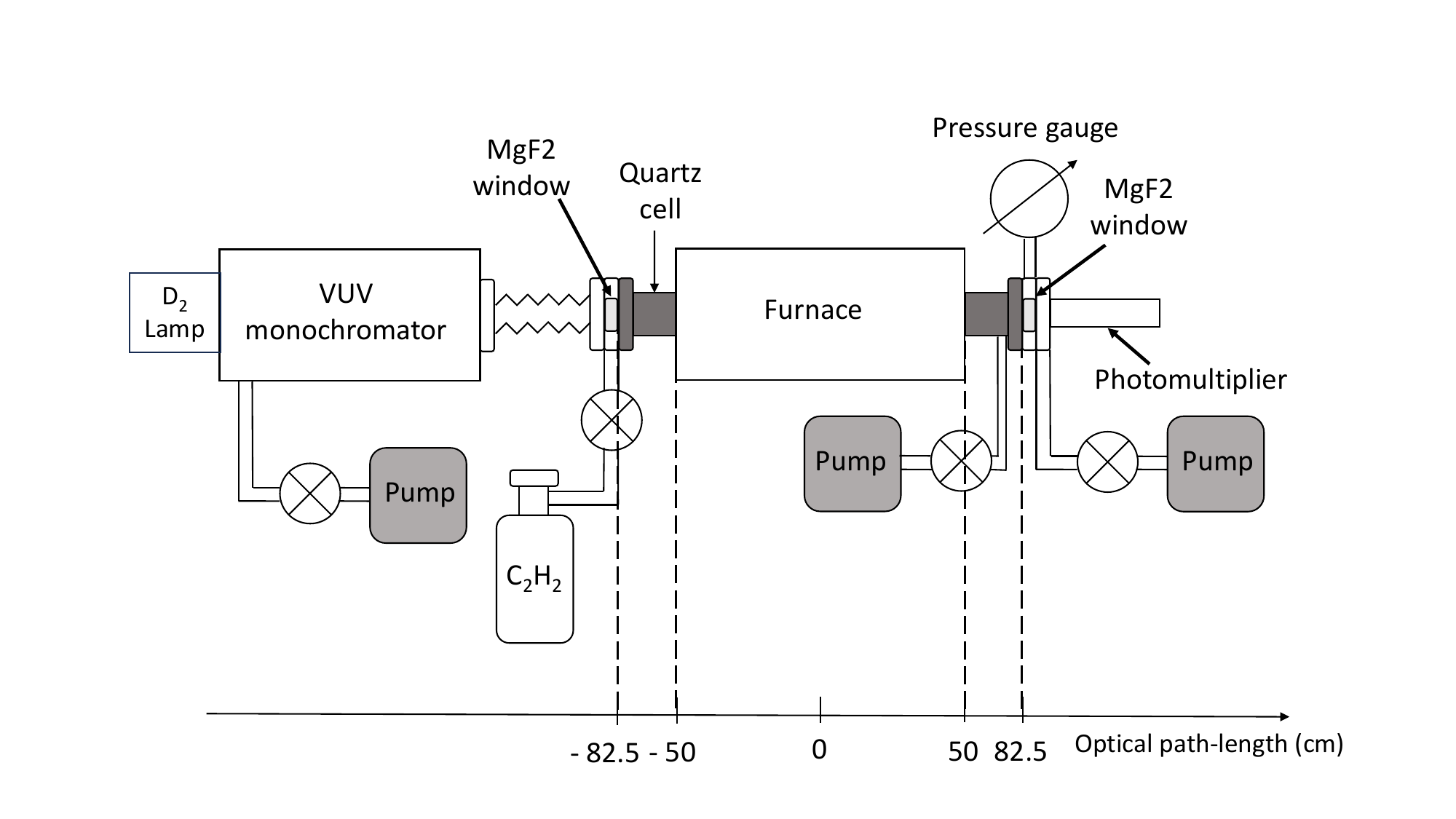}
    \caption {Experimental setup used at LISA to measure the VUV spectra of acetylene.}
    \label{Figure_1}
\end{figure*}

The absorption spectra of gaseous acetylene were measured at high temperatures using a new custom-made VUV spectroscopy platform developed at the Laboratoire Interuniversitaire des Systèmes Atmosphériques (LISA, France). A scheme of the experimental setup is presented in Fig. \ref{Figure_1}.

The setup consists of a high-temperature absorption cell composed of a quartz tube sealed at each extremity with magnesium fluoride (MgF$_2$) windows (16~mm in diameter) mounted on stainless-steel flanges. The optical pathlength of the cell was 165~cm. The cell was installed in a furnace that can warm up to 1373~K. The temperature of the cell was continuously measured with three type-K thermocouples equally spaced along its length. Before each measurement at a given temperature, the cell was pumped using a turbomolecular pump and heated up to the studied temperature for 24 hours to desorb and evacuate gases. At ambient temperature, the background pressure was 10$^{-7}$~mbar. For this study, spectra were measured at ambient temperature (i.e., 296~K) as well as at 373, 473, 573, 673, and 773~K. We limited our measurements to a maximum temperature of 773~K due to rapid thermal decomposition of C$_{2}$H$_{2}$ observed at elevated temperatures, probably enhanced by catalytic processes onto the surface of the cell (see discussion in Sect. \ref{subsec:Evolution of C2H2 absorption cross section with the temperature}). 

The VUV spectra of C$_{2}$H$_{2}$ were measured from 115 to 230~nm using a McPherson 225 ultrahigh vacuum (UHV) monochromator (1~m focal length) pumped down to 8 $\times$ 10$^{-8}$ mbar using an ionic pump to prevent absorption by atmospheric molecules (nitrogen, N$_2$, oxygen, O$_2$, etc.). The limits of the wavelength range are due, respectively, to the MgF$_2$ windows cut-on and the first order diffraction grating’s limit. At the entrance of the monochromator, a 30~W deuterium (D$_2$) lamp produces a polychromatic VUV-UV beam directed toward the diffraction grating (1200 slits/mm, blazed at 140~nm), which reflects and disperses the light to obtain a quasi-monochromatic beam at the exit slit. Then, the beam passes through the cell and the photons are detected by a Hamamatsu R6836 photomultiplier tube producing a current proportional to the photon flux. Finally, the current is measured and digitalized using a Keithley 6485 pico-amperemeter. The entrance and exit of the monochromator are equipped with slits and occulters. The height of occulters can be adjusted from 2 to 20~mm and the width of slits from 0.005 to 2~mm. For this study, entry and exit occulter heights were set respectively to 6 and 4~mm, while both the entry and exit slits’ widths were set to 40~$\muup$m. With these slits’ width, the spectral resolution of the monochromator, which corresponds to the full width at half maximum of a spectral band, is 0.034~nm. However, this spectral resolution was further limited to $\sim$ 0.1~nm by the chosen sampling interval (one point every 0.03~nm). In the case of measurements using the LISA setup, a background spectrum with an empty cell was recorded at the beginning of each measurement, followed by the acquisition of various absorption spectra for C$_{2}$H$_{2}$. The background spectrum was obtained after allowing the D$_2$ lamp to warm up for at least one hour. We noted that during the first hour after being turned on, the emission intensity of the D$_2$ lamp could fluctuate by a few percent. After this initial warm-up period, the emission intensity stabilizes sufficiently, contributing only slightly to the uncertainty, which is accounted for in the measured spectra.

As the acetylene cross section spans a variation of five orders of magnitude in the studied spectral domain (see Fig. \ref{Figure_2}), it was necessary to employ a wide range of pressures to accurately record the smallest absorptions and prevent any saturation of the strongest absorption bands. To facilitate our measurements, we divided the full wavelength range into ten overlapping segments of a few nanometers each. For each measurement, gaseous acetylene (Air Liquide, purity > 99.5\%) was introduced into the cell at the desired pressure (ranging from 10$^{-3}$ to 80~mbar) and all the considered segments were measured at least three times at different pressures. In total, fifty spectra were measured at each temperature. The absolute pressure of C$_{2}$H$_{2}$ in the cell was measured with two capacitance gauges ranging from 10$^{-4}$ to 1~mbar and from 0.1 to 1000~mbar, respectively. As was discussed in \citet{Benilanetal.1995}, the presence of acetone traces inside the gas cylinder (used as a stabilizer) can affect the spectra of C$_{2}$H$_{2}$ recorded in certain part of the VUV domain ($\sim$185 - 196~nm) where acetone cross sections \citep{Nobreetal.2008} are large compared to acetylene ones. To prevent this issue when measuring the corresponding spectral domain, acetone present in the injected gas was removed by circulating the gas before injection into a cold trap made of a stainless-steel coil immersed in a slush of liquid nitrogen and acetone at $\sim$183~K.

Additional spectra of C$_{2}$H$_{2}$ were measured at 296, 573, and 773~K at synchrotron SOLEIL (proposal 20150691, PI: Venot) on the DESIRS (Dichroïsme Et Spectroscopie par Interaction avec le Rayonnement Synchrotron) VUV beamline, whose wavelength range extends from 30 to 250~nm \citep{Nahonetal.2012}. Although mostly similar, the method employed at SOLEIL presented a few differences. Firstly, measurements were made using a metallic cell made of Kanthal material (instead of quartz) with a shorter optical pathlength of 146~cm. This cell was used previously to measure CO$_2$ spectra  \citep{Venotetal.2013, venotetal2018}. Second, DESIRS beamline is equipped with a Fourier Transform Spectrometer (FTS) that allows to record spectra at higher resolution (resolving power up to $\sim$10$^6$) and with a high absolute wavelength accuracy of 1 $\times$ 10$^{-7}$ \citep{deOliveiraetal.2011}. At 296~K, spectra were measured from 160.82 to 196.08~nm with a resolution of 0.0134~nm and from 196.08 to 235.43~nm with a resolution of 0.0034~nm. At 573 and 773~K, spectra were measured from 154.37 to 169.49~nm with a resolution of 0.0537~nm and from 169.49 to 203.75~nm with a resolution of 0.0134~nm.

\subsection{Data processing} \label{subsec:Data processing}
\subsubsection{Calculation of the absorption cross section} \label{subsubsec:Calculation of the absorption cross section}

The absolute absorption cross section of C$_{2}$H$_{2}$ was calculated from the measured spectra using the Beer-Lambert law:

\begin{equation}
    \sigma(\lambda,T) = \frac{1}{nL} \times \ln{(\frac{I}{I_0})},
    \label{Eq.1}
\end{equation}
where $\sigma$($\lambda$, T) is the absorption cross section (cm$^{2}$) at a given wavelength, \emph{$\lambda$}, and temperature, \emph{T}, \emph{L} is the optical pathlength (cm), \emph{n} the volume density of the gas in the cell (cm$^{-3}$), \emph{I$_0$} the intensity of the light transmitted through an empty cell, and \emph{I} the intensity of the light transmitted through the cell containing a density \emph{n} of gas. Considering the gas as a perfect gas, the density of the gas in the cell can be calculated using the equation \emph{n=P/$\rm{k_B}$T}, where \emph{P} is the pressure of C$_{2}$H$_{2}$ inside the cell (Pa), \emph{T} the temperature (K), and $\rm{k_B}$ the Boltzmann constant.

Prior calculating the absorption cross section, it is necessary to eliminate data points from each measured spectrum that are too noisy (for low pressure) or saturated (for high pressure) by applying two filters. The first filter eliminates noisy points with a transmission higher than a fixed value (typically 80\%). The second filter removes saturated points whose optical depth does not increase linearly with the gas pressure. Then, for the ten segments and each pressure, the absorption cross section was calculated using Eq. \ref{Eq.1} and a mean value was determined. Finally, the ten spectra obtained were concatenated to reconstruct the full spectrum ranging from 115 to 230~nm.

\subsubsection{Temperature and wavelength calibrations} \label{subsubsec:Temperature and wavelength calibrations}

The temperature gradient along the entire absorption cell was quantified in our previous studies by \citet{Venotetal.2013, venotetal2018}, which examined the effect of high temperature on the VUV cross sections of CO$_2$ molecules. The temperature reaches its peak at the cell center, equal to the set oven temperature, and then symmetrically decreases toward near-ambient at the cell edges. Near the center, the temperature remains maximal over a length of about 35~cm. The transition section, where the temperature drops from maximum to ambient, spans approximately 25~cm in length, while the zones near ambient temperature are roughly 22.5~cm. As the absorption is dominated by the gas at high temperature, to calculate the gas density and then the absorption cross section, we decided to follow the same assumption as in \citet{venotetal2018} and to consider that the temperature of the gas is equal to the maximum temperature (T$_{\mathrm{max}}$) all along the optical pathlength. The measurement uncertainties in the absorption cross sections of C$_{2}$H$_{2}$ were determined using a root-sum-of-squares analysis, as was described by \citet{Meyn, ESsebbar2014}. According to Eq. \ref{Eq.1}, the uncertainty contributions considered from different sources, include concentration (i.e., impurities) $\Delta$n/n, optical path length $\Delta$l/l, temperature $\Delta$T/T, pressure $\Delta$P/P, and measured absorbance $\Delta$A/A. The concentration uncertainty was estimated to be under 0.5\%, primarily due to a maximum of 0.5\% impurities, possibly acetone, present in the acetylene gas cylinder \citep{Benilanetal.1995}. The uncertainty in the optical path length was calculated to be less than 1\%. This was determined by comparing absorption data measured at room temperature in two absorption cells with optical path lengths of 165~cm and 146~cm at LISA and SOLEIL, respectively. The pressure measurement, considering both the manufacturing accuracy of the Baratron capacitance manometers and minor pressure variations during recording, particularly at high resolution, has an estimated relative uncertainty of 0.5\%. The error in the absorbance ranged from 1 to 2\%, which also accounts for the slight variations in the background intensity of the D$_2$ lamp. At ambient temperature, the temperature uncertainty in the cell yields a relative uncertainty of $\Delta$T/T $\sim$ 0.5\%. Overall, uncertainty in the measured cross sections at ambient temperature is estimated to be below 6\%, accounting for all sources of error.
However, at high temperatures, temperature gradients become a significant source of uncertainty among all contributing factors, as the temperature is not uniform along the absorption cell. This temperature variation induces changes in gas density along the cell while pressure remains constant. Subsequently, the uncertainty in the absorption measured at a given maximum temperature, T$_{\mathrm{max}}$, largely depends on the variation in the absorption cross section of C$_{2}$H$_{2}$ between the cooler and warmer regions of the cell. More recently, \citet{poveda:tel-04718446} investigated the effect of temperature gradients on the measured absorption cross sections of CO. This study demonstrated that most of the absorption from the hot bands occurs within a $\pm$ 40~cm region at the center of the cell, where the temperature is highest, promoting the formation of hot bands. Consequently, the temperature gradient introduces only a minor additional uncertainty to the measured cross section for these hot bands. However, the results of our study provide limited information on the cold bands. Assuming equal lengths for the cold and warm sections of the cell, the column density would vary as 1/T. With a gas temperature roughly twice as high at the center compared to the extremities of the cell, the column density would be approximately twice as high in the colder section. This could lead to an underestimation of the absorption cross section at high temperature by several tens of percent. Further work is needed to more precisely evaluate the uncertainty introduced by the temperature gradient.

The wavelength calibration of the spectra measured at LISA was carried out on the basis of the spectra measured at SOLEIL because of the high absolute wavelength accuracy of the spectra measured with the FTS of the DESIRS beamline \citep{deOliveiraetal.2011}. For the comparison and calibration purpose of the spectra, it was necessary to first degrade the SOLEIL spectra to the lower resolution used at LISA (0.1~nm).

\section{Results and discussions} \label{sec:Results and discussions}
\subsection{Absorption cross sections of C$_{2}$H$_{2}$ at 296~K} \label{subsec:Absorption cross sections of C2H2 at 296 K}
\subsubsection{New data obtained at LISA and SOLEIL facilities} \label{subsubsec:New data obtained at LISA and SOLEIL facilities}

\begin{figure}[h]
    \centering
    \includegraphics [trim=40 0 80 0,clip,width=\hsize] {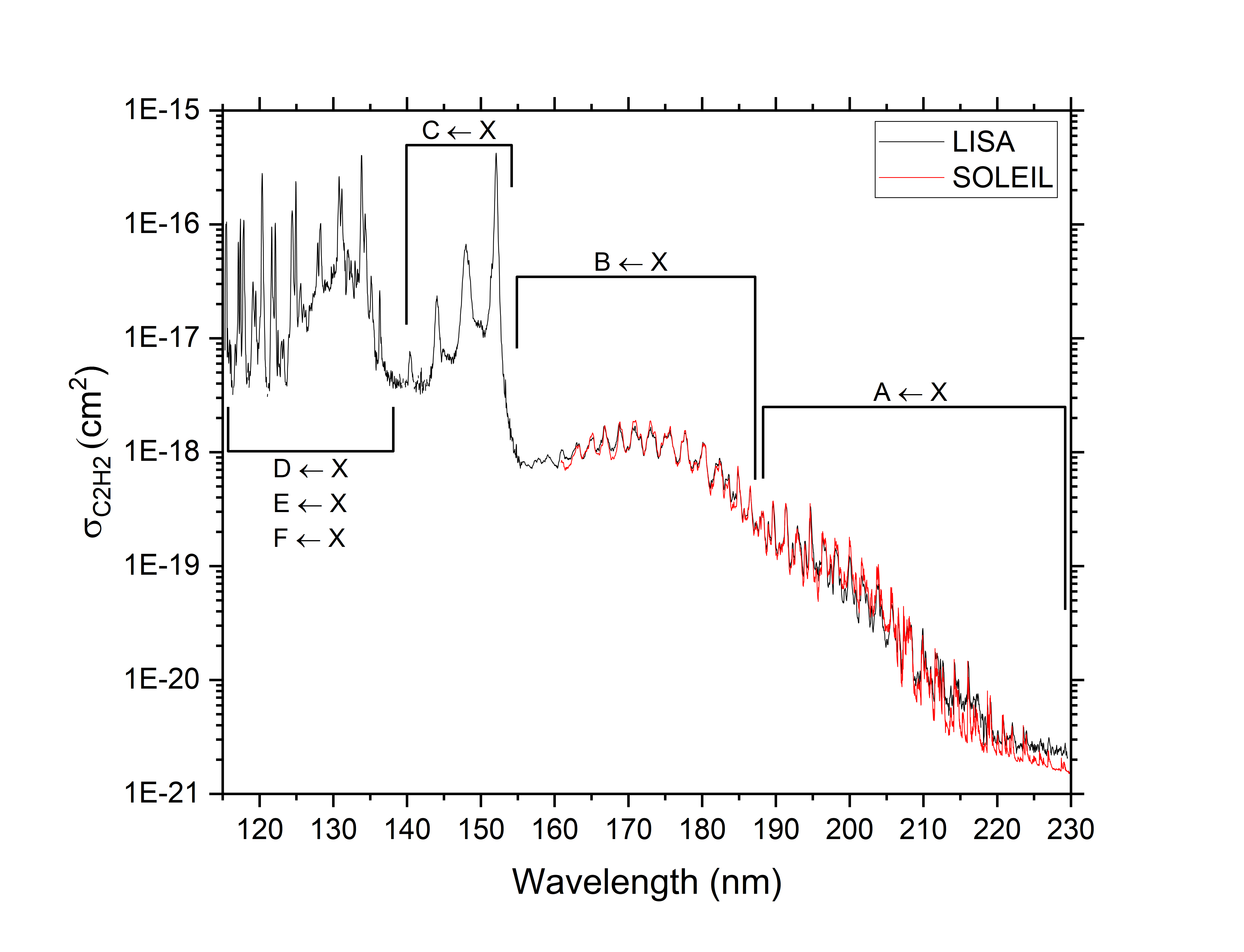}
    \caption {Absorption cross section of C$_{2}$H$_{2}$ determined from 115 to 230~nm at LISA (black) and from 160.8 to 230~nm at SOLEIL with a resolution of 0.1~nm (red). Electronic transitions associated with observed absorption bands are indicated.}
    \label{Figure_2}
\end{figure}

We started our study by measuring the absorption cross section of C$_{2}$H$_{2}$ at 296~K. Data obtained at LISA from 115 to 230~nm as well as the ones obtained at SOLEIL (after degrading the resolution of the spectrum to 0.1~nm) from 160.82 to 230~nm are presented in Fig. \ref{Figure_2}. The spectra present two components: a continuum and different band systems superimposed on it. Band’s features observed from 185 to 230~nm and from 153 to 185~nm are attributed, respectively, to the A$^1$A$_u$ $\leftarrow$ X$^1\Sigma_g^+$ and B$^1$B$_u$ $\leftarrow$ X$^1\Sigma_g^+$ electronic transitions from the ground state X to the two first excited states A and B \citep{FooandInes.1973, Watsonetal.1982, VanCraenetal.1985, VanCraenetal.1986, Wuetal.1989, Chenetal.1991, Smithetal.1991}. Sharp and intense absorption bands from 140 to 155~nm are attributed to the C$^1\Pi_u$ $\leftarrow$ X$^1\Sigma_g^+$ electronic transition to the C Rydberg state \citep{Herzberg.1966, GedankenandSchnepp.1976}. Finally, absorption bands observed at wavelengths shorter than 140~nm are attributed to the D$^1\Sigma_u^+$ $\leftarrow$ X$^1\Sigma_g^+$, E$^1\Pi_u$ $\leftarrow$ X$^1\Sigma_g^+$, and F$^1\Pi_u$ $\leftarrow$ X$^1\Sigma_g^+$ electronic transitions involving the D and F Rydberg states as well as the E valence state \citep{Wuetal.2001, Boyeetal.2004, Chengetal.2011}. In general, the two sets of data agree in terms of absolute values of the intensity, but we observed some differences in the continuum level of the A $\leftarrow$ X system. From 195 to 205 nm, the SOLEIL values are higher than LISA ones, while from 210 to 230 nm, the continuum level is lower in the SOLEIL spectrum than in the LISA one. Nevertheless, in both cases, there is $\sim$25\% of differences between the continuum values. In addition, we also observed in this 195-205~nm region that the absorption bands in the LISA spectrum are less intense than those in the SOLEIL spectrum. The origin of these differences is difficult to establish, but we can formulate some hypothesis. Firstly, we observe that the region where the intensity of the SOLEIL spectrum is more intense than the LISA one (195 to 205~nm) corresponds to the region for which the acetone traces present inside the gas cylinder absorbs and, thus may affects the calculation of the absorption cross section of C$_{2}$H$_{2}$ \citep{Benilanetal.1995}. Therefore, our result may suggest than the acetone have not been totally trapped in one of the experiments, resulting in an additional absorption in this spectral region. For the 210-230~nm region, we note that the pressures used at SOLEIL and LISA are not identical. Higher pressures were used for some of the SOLEIL spectra compared to those of LISA. If the pressures used at LISA were too low, it may lead to a higher noise level on the continuum, resulting in a higher uncertainty on the intensity of the continuum in this region. The high-resolution spectrum obtained at SOLEIL, presented in Fig. \ref{Figure_5}, reveals that at longer wavelengths (i.e., $\lambda$ > 185~nm), corresponding to the C$_{2}$H$_{2}$ (A $\leftarrow$ X) band system, rotational structures are well resolved. However, at shorter wavelengths below 185~nm, predissociation causes a broadening and a decrease in absorption intensity, resulting in bands without visible rotational features. In the lower-resolution LISA spectrum, only the predissociation-broadened bands are resolved and distinguishable, while the rotational structures at longer wavelengths are no longer resolved.

\subsubsection{Comparison with data available in the literature} \label{subsubsec:Comparison with data available in the literature}

\begin{figure}[h]
    \centering
    \includegraphics [trim=50 0 80 0,clip,width=\hsize] {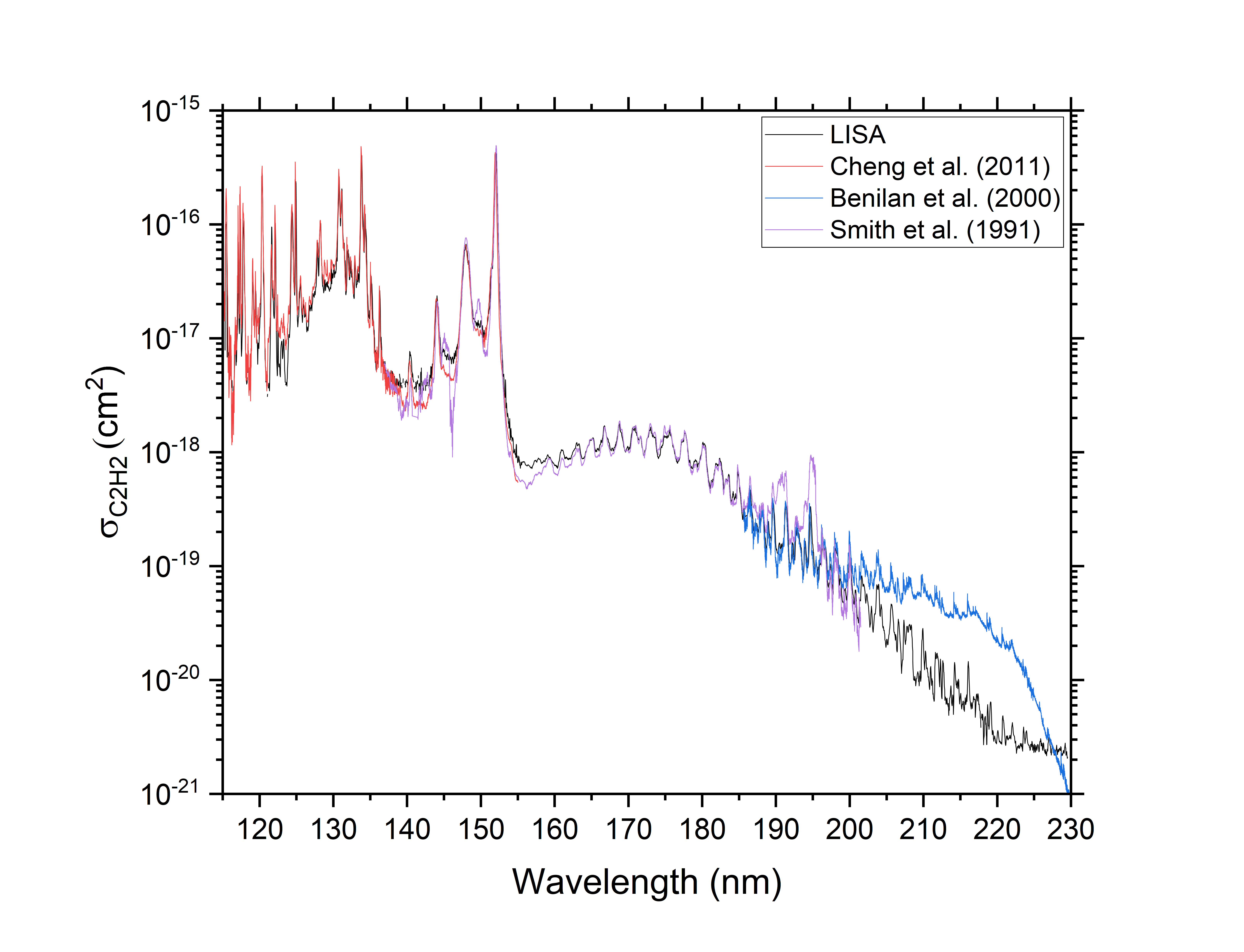}
    \caption {Comparison of absorption cross of C$_{2}$H$_{2}$ section measured at ambient temperature from 115 to 230~nm at LISA (black) with the ones measured by \citet{Chengetal.2011} (red), \citet{Smithetal.1991} (purple), and \citet{Benilanetal.2000} (blue).}
    \label{Figure_3}
\end{figure}

We compare in Fig. \ref{Figure_3} the absorption cross section measured at ambient temperature with data available in the literature \citep{Smithetal.1991, Benilanetal.2000, Chengetal.2011} and which cover the whole wavelength range of our study (i.e., 115-230~nm). 
For 115 < $\lambda$ < 185~nm, our data generally agree with the ones measured by \citet{Smithetal.1991, Chengetal.2011} despite observing some differences in the intensity of the baseline, especially for $\lambda$ < 140~nm as well as around, 145, and 155~nm. We note that $\sim$155~nm, the continuum in LISA spectrum does not decrease as much as in \citet{Smithetal.1991, Chengetal.2011} toward longer wavelengths. As was mentioned previously, these differences could be explained by using lower pressures in this spectral range, impacting the accurate determination of the continuum intensity. Additionally, the LISA spectra exhibit a slight noise around 155~nm due to numerous emission bands from the D$_2$ lamp (used as light source), particularly the large emission from 155 to 165~nm features of molecular D$_2$ (B$^1\Sigma_u$ $\leftarrow$ X$^1\Sigma_g$). Yet, these bands are not resolved with the settings used for our measurements. In addition, the intensity of those bands may vary of a few percent over time due to variations in the lamp’s temperature. Subsequently, this contributes to a slight increase in the noise level and uncertainty of the intensity of the continuum in this region. 

From 185 to 195~nm, our data agree with \citet{Benilanetal.2000} but we observe important differences compared to the data from \citet{Smithetal.1991}, which exhibit two large absorption bands that are absent from the other spectra. As was described in Section \ref{subsec:Measurements of acetylene UV absorption spectra}, we suppose that these bands are due to the contamination of C$_{2}$H$_{2}$ by acetone \citep{Benilanetal.1995}, which was not trapped in the study by \citet{Smithetal.1991} contrary to our measurements and those of \citet{Benilanetal.2000}.
Beyond 195~nm, we observed a good agreement between our data and those of \citet{Smithetal.1991}, which extends up to 200~nm. However, we observed important differences (up to one order of magnitude) with the data measured by \citet{Benilanetal.2000} from 195 to 230~nm. Indeed, while in our spectrum the intensity of the baseline continues to decrease exponentially from 195 to 230~nm, the intensity of the spectrum measured by these authors decreases much slowly up to 220~nm before decreasing drastically up to 230~nm. Although it is not possible to identify with certainty the origin of these differences, the higher absorption cross section determined by \citet{Benilanetal.2000} would suggest that the gas used by these authors could have been contaminated with another species that absorb at these wavelengths, leading to a bias in the determination of the absorption cross section of C$_{2}$H$_{2}$ in this spectral region.

\subsection{Evolution of C$_{2}$H$_{2}$ absorption cross section with the temperature} \label{subsec:Evolution of C2H2 absorption cross section with the temperature}

\begin{figure}[h]
    \centering
        \includegraphics[trim=40 0 80 40,clip, width=0.5\textwidth]{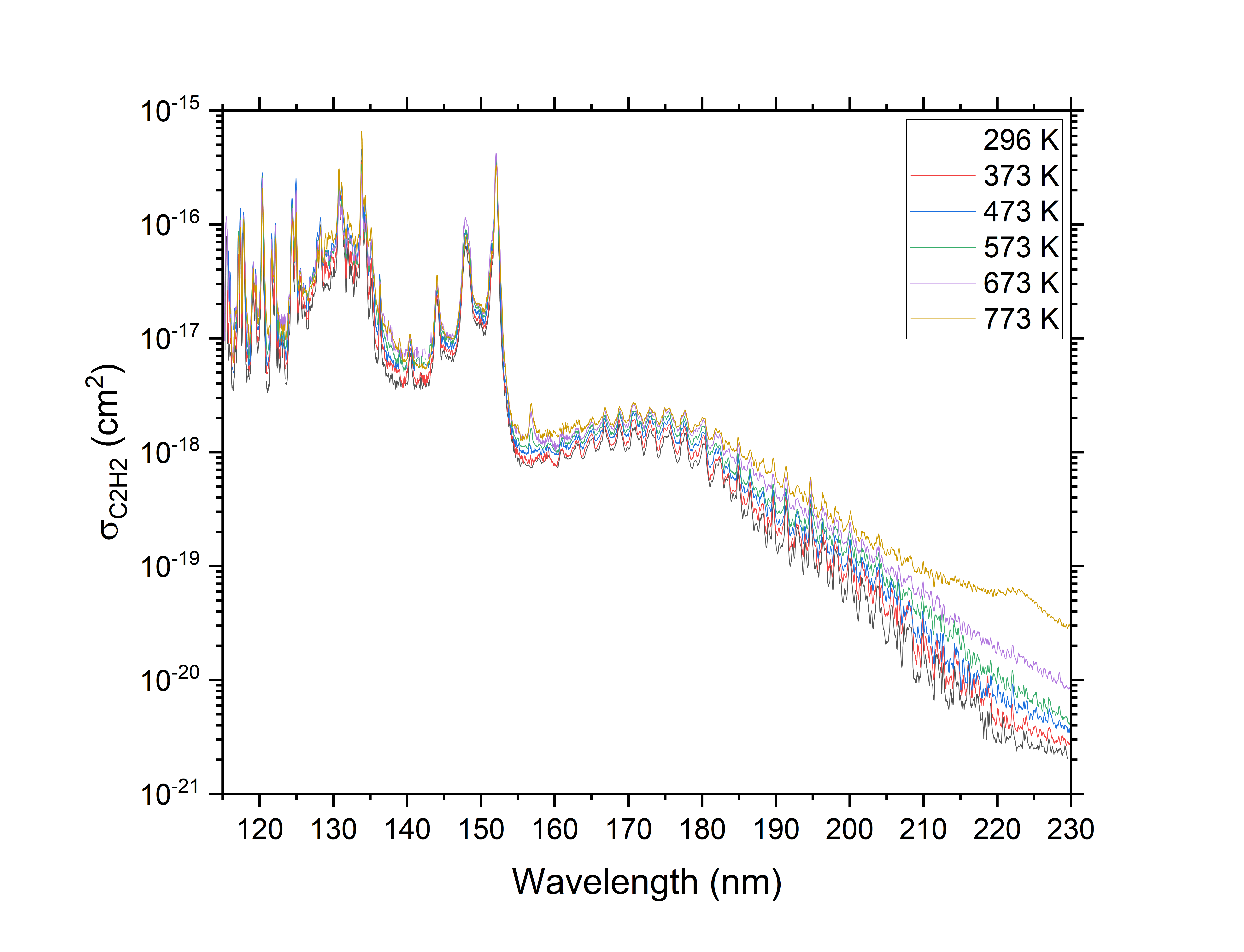}
        \includegraphics[trim=40 0 80 50,clip, width=0.5\textwidth]{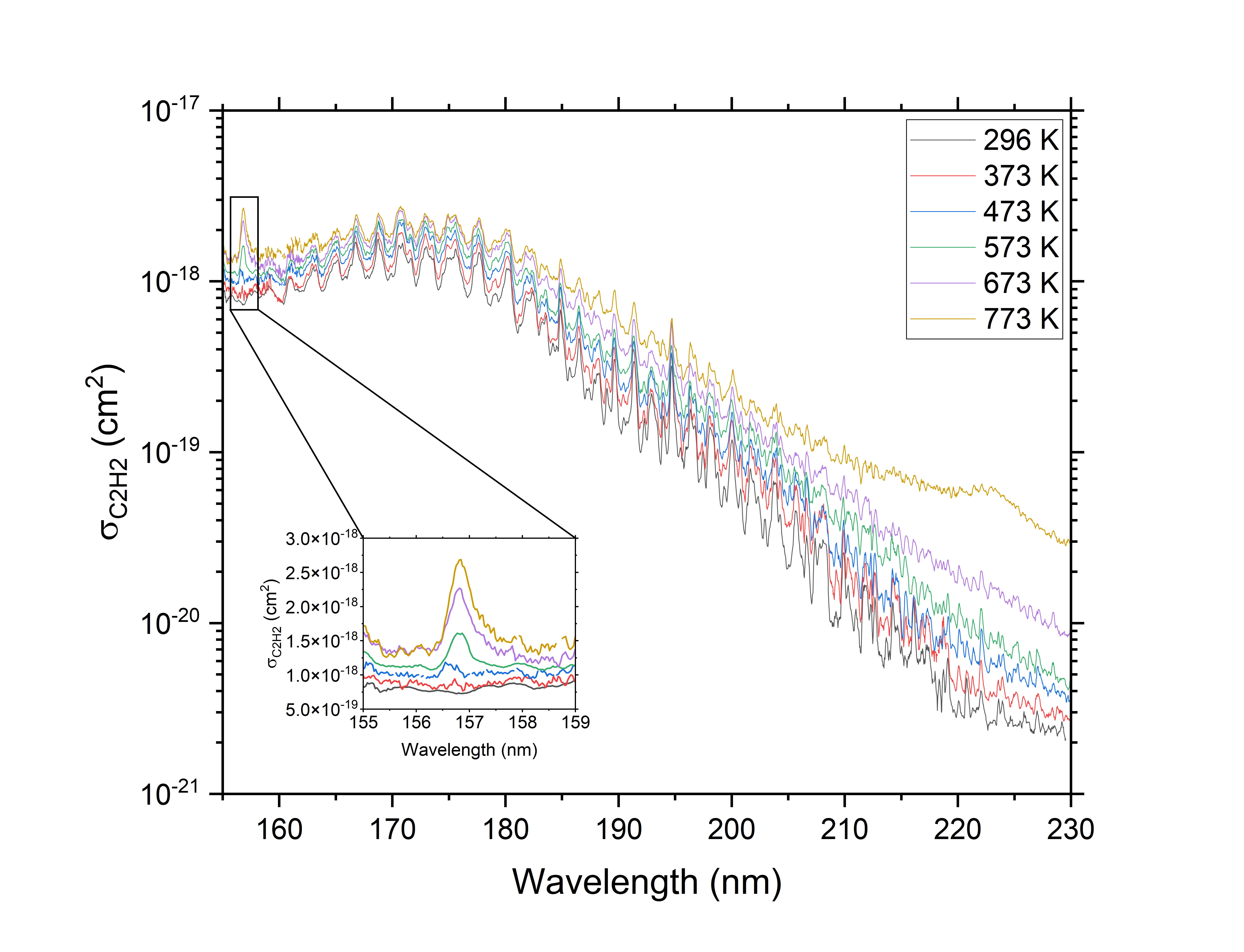}
    \caption {Absorption cross section of C$_{2}$H$_{2}$ determined at LISA at 296~K (black), 373~K (red), 473~K (blue), 573~K (green), 673~K (violet), and 773~K (yellow). Top: Spectra covering the whole studied spectral range from 115 to 230~nm. Bottom: Same spectra with a focus on the 155-230~nm range and on the 155-159~nm range, as is shown in the inset.} 
    \label{Figure_4}
\end{figure}

Following the measurements at 296~K, we determined the absorption cross section of C$_{2}$H$_{2}$ at 373, 473, 573, 673, and 773~K. Figure \ref{Figure_4} presents the evolution of the absolute absorption cross section of C$_{2}$H$_{2}$ measured from 115 to 230~nm at LISA as a function of temperature. In general, we observe an increase in the intensity of the continuum and a decrease in the intensity of the absorption bands as the gas temperature increase. This results in a net increase in the absolute absorption cross section of C$_{2}$H$_{2}$ with temperature. In the 160 to 230~nm range, the absorption cross section increases linearly with the temperature. However, we observe at 773~K a more important increase in the absorption cross section for wavelengths above 210~nm compared to the other temperatures, with the appearance of a bump centered at 222.5~nm. We were not able to determine the absorption cross section of C$_{2}$H$_{2}$ at higher temperature because we observed a fast thermal degradation at 873~K. As a result, the gas density inside the cell was not constant over time and could not be determined based on the measured gas pressure. Therefore, we could not calculate the absorption cross section of C$_{2}$H$_{2}$ at 873~K based on the spectra measured experimentally, and thus did not attempt to go to even higher temperatures. Due to the role of C$_{2}$H$_{2}$ in the formation of other hydrocarbons and soot in combustion systems, the thermal decomposition of C$_{2}$H$_{2}$ has been extensively studied. It has been observed that in static conditions (like the ones used in our experiments), acetylene pyrolysis can occur at temperatures lower than 1000~K \citep{Duranetal.1988, Colketetal.1989, Kernetal.1991, Benson1992, Zadoretal.2017, Liuetal.2021}. This agrees with our observations, where C$_{2}$H$_{2}$ thermal degradation may be further enhanced by the walls of the absorption cell. For temperatures lower than 1500~K, the mechanism responsible for the pyrolysis of acetylene has not been firmly established and different initiation reactions have been proposed, such as acetylene dimerization or self-reaction \citep{Benson1992, Zadoretal.2017}. Nevertheless, the thermal degradation of C$_{2}$H$_{2}$ can lead to the formation of new molecules such as polyacetylenes \citep{Aghsaeeetal.2014}, which may affect the absorbance measured experimentally resulting in an overestimation of the absorption cross section. This has been shown by \citet{Zabetietal.2017} who determined experimentally the absorption cross section of C$_{2}$H$_{2}$ from 200 to 400~nm for temperatures up to 1500~K using a large shock tube facility. They reported values five times lower than those previously determined by \citet{Vattulainenetal.1997}. They concluded that the measurements of \citet{Vattulainenetal.1997}, conducted in static systems with longer residence times (a few minutes), were affected by the pyrolysis of C$_{2}$H$_{2}$. In contrast, in the study reported by \citet{Zabetietal.2017}, the effect of the thermal degradation of C$_{2}$H$_{2}$ was negligible, due to the much shorter residence time (a few $\muup$s). These two observations suggest that in the LISA experiments, the significant increase in the absorption cross section observed at 773~K above 210~nm could be attributed to the thermal degradation of C$_{2}$H$_{2}$ at this temperature and the formation of new absorbing species. Additional work will be needed to confirm this hypothesis by measuring the evolution of the gas composition inside the cell as a function of time at 773~K. 

In addition to these general variations, we observe at 473~K and beyond that a new absorption band appears at $\sim$156.7~nm (see inset in Fig. \ref{Figure_4}-bottom), with an intensity that increases with the temperature. We identify this band as a hot band apparently originating from the $\nu_2$ vibrational mode ($\nu_2$ = 1973.8~cm$^{-1}$). Indeed, the origin of this transition is at 65860~cm$^{-1}$ \citep{GedankenandSchnepp.1976} while the hot band is at 63816~cm$^{-1}$. Therefore, the two vibrational levels are separated by 1973~cm$^{-1}$, which corresponds to the vibrational mode $\nu_2$.

\begin{figure}[h]
    \centering
        \includegraphics[trim=40 0 80 40,clip, width=0.5\textwidth]{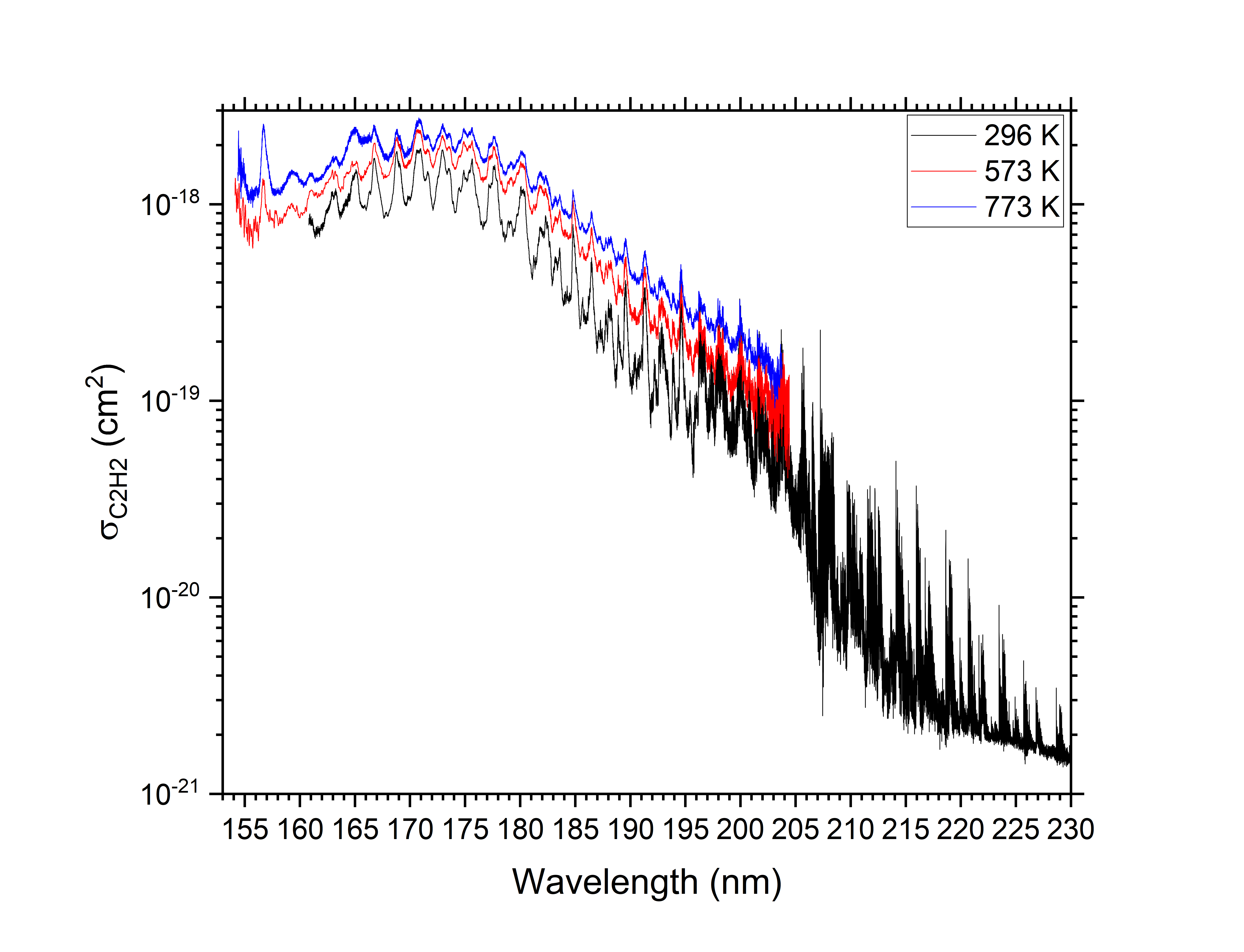}
        \includegraphics[trim=40 0 80 40,clip, width=0.5\textwidth]{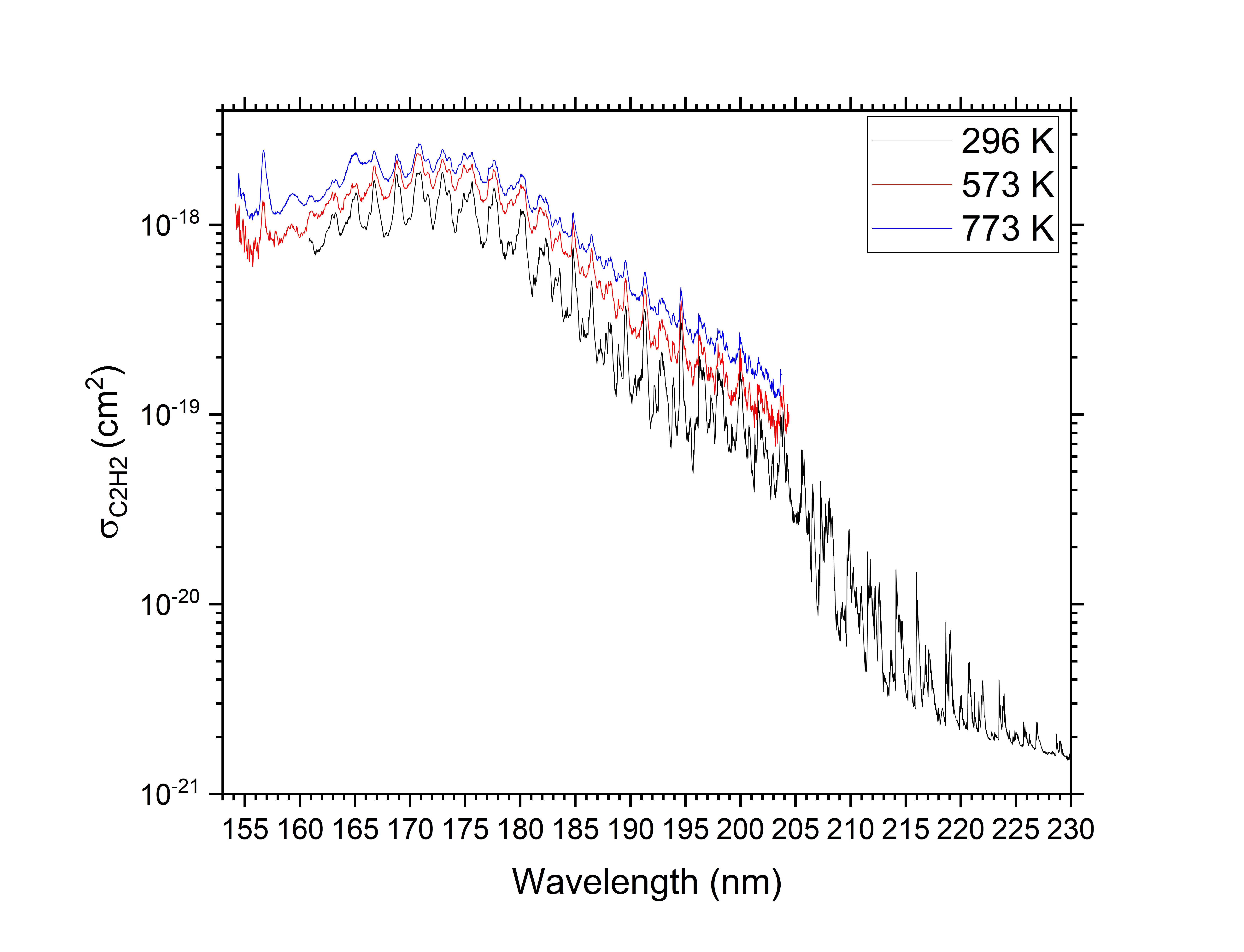}
    \caption {Absorption cross section of C$_{2}$H$_{2}$ measured at synchrotron SOLEIL from $\sim$154 to 230~nm at 296~K (black), 573~K (red), and 773~K (blue). Top: Spectra measured at SOLEIL with resolutions ranging from 0.0134 to 0.0034~nm. Bottom: Same spectra after a degradation of their resolution to 0.1 ~nm (the one used at LISA).}
    \label{Figure_5}
\end{figure}

Figure \ref{Figure_5} presents the absorption cross sections of C$_{2}$H$_{2}$ obtained at SOLEIL at 296, 573, and 773~K at high resolution (top) and after degrading the spectral resolution alike that of LISA spectra. Similar to LISA data, we observe an increase in the continuum intensity and a decrease in the intensity of the absorption bands when the temperature increases. Moreover, we observed the apparition of the same “hot band” at $\sim$156.7~nm in the spectra recorded at 573 and 773~K. However, the spectra recorded at high temperatures at SOLEIL do not extend beyond 205~nm, which prevents the study of a large part of the A $\leftarrow$ X band system where we observed the most significant variation in the absorption cross section with temperature at LISA (see Fig.\ref{Figure_4}).

\begin{figure}[h]
    \centering
        \includegraphics[trim=40 0 80 40,clip, width=0.5\textwidth]{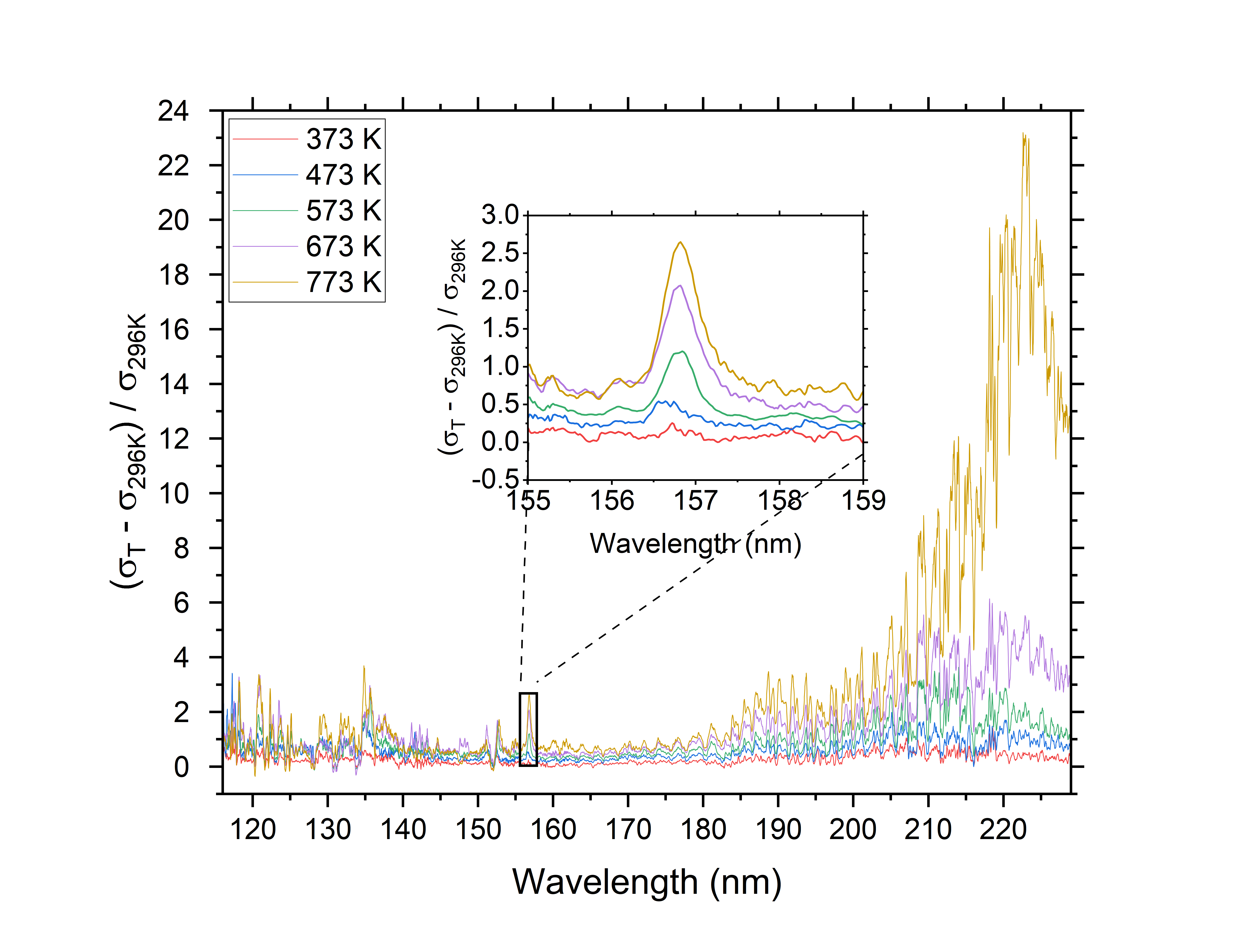}
        \includegraphics[trim=40 0 80 40,clip, width=0.5\textwidth]{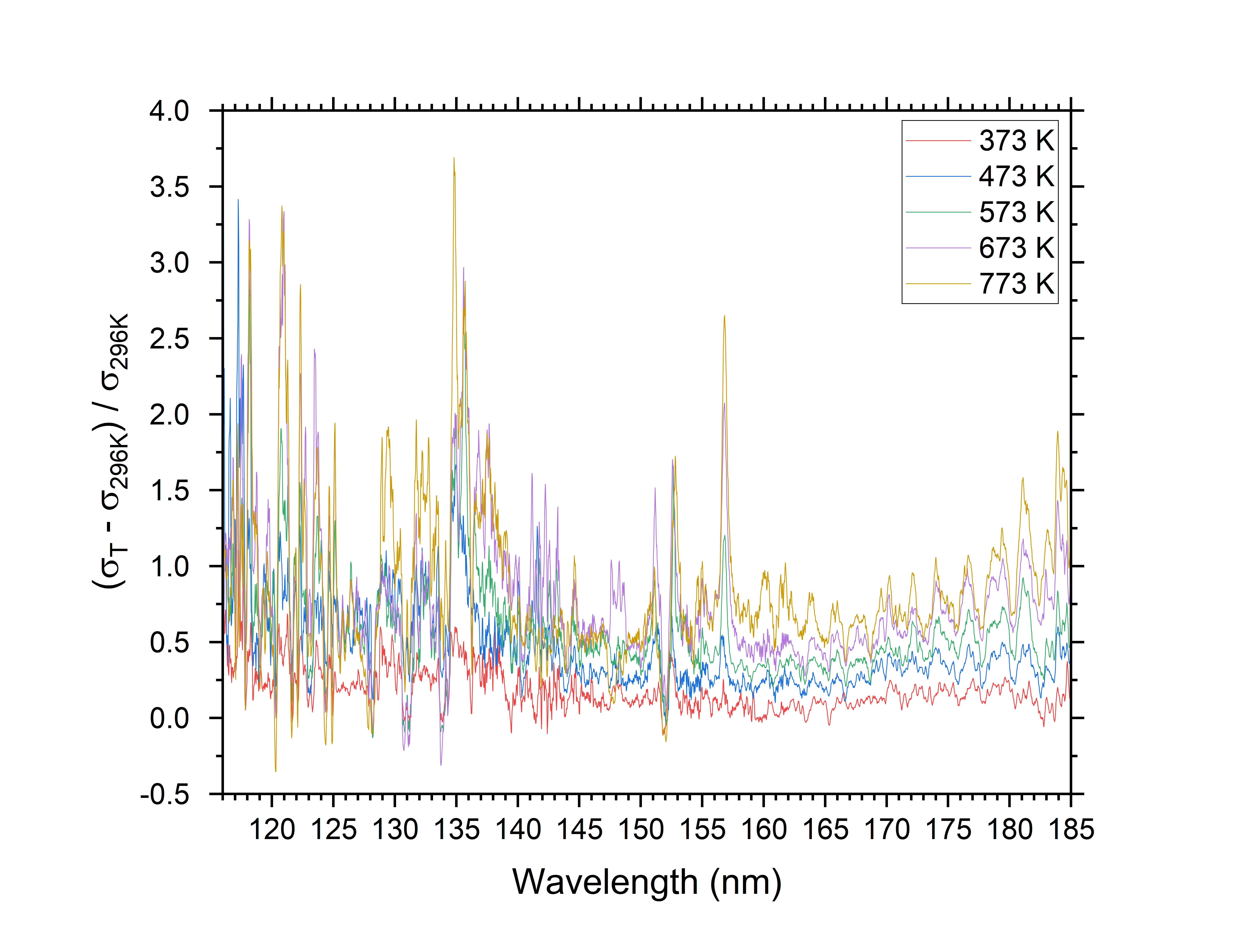}
    \caption {Factor of changes, \emph{F$_T$}, in the absorption cross section of C$_{2}$H$_{2}$ between high temperatures, T (K), and the ambient temperature (296~K) for LISA measurements. Top: \emph{F$_T$} calculated for the whole studied spectral range (116 to 229~nm) along with an inset of 155-159~nm region where a hot band is visible. Bottom: Same spectra showing only the 116 to 185~nm region. }
    \label{Figure_6}
\end{figure}

To go further in the quantification of the change in the absorption cross section with the temperature, we calculated the factor of changes of the cross section, defined by \citet{Wuetal.2001}.

\begin{equation}
    F_T = \frac{\sigma_T - \sigma_{296K}}{\sigma_{296K}},
    \label{Eq.2}
\end{equation}
where \emph{F$_T$} is the factor of changes of cross section between the temperature, \emph{$T$}, and our reference at 296~K, \emph{$\sigma_T$} is the absorption cross section at \emph{T}, and \emph{$\sigma_{\rm{296K}}$} is the absorption cross section at 296~K. A positive value corresponds to an increase in the cross section at \emph{T} comparatively to 296~K, while a negative value corresponds to a decrease. The calculation of \emph{F$_T$} for the absorption cross section obtained at LISA (116 to 229~nm) at the different temperatures is presented in Fig. \ref{Figure_6}. From 116 to 185~nm, which corresponds to the B $\leftarrow$ X transition at 185~nm, the continuum increase is relatively constant on the whole wavelengths range at a given temperature, while the intensity of the absorption bands decreases. At 373~K, the cross section of C$_{2}$H$_{2}$ increases with an average \emph{F$_T$} of 0.2, similarly to the value found by \citet{Wuetal.2001} from 120 to 140~nm at 370~K. For the continuum, this factor increases with temperature up to an average of 0.5 at 773~K, while variations in certain bands can be more significant, up to a factor of 3.5. Beyond 185~nm (see Fig. \ref{Figure_6} top), \emph{F$_T$} increases sharply toward longer wavelengths until it reaches 195~nm, where it then decreases. Then, \emph{F$_T$} rises sharply again along the C$_{2}$H$_{2}$ (A $\leftarrow$ X) bands system until 215 nm, where another decrease is observed. Finally, \emph{F$_T$} increases up to 220~nm before decreasing again until 230~nm.  

\begin{figure}[h]
    \centering
    \includegraphics [width=\hsize] {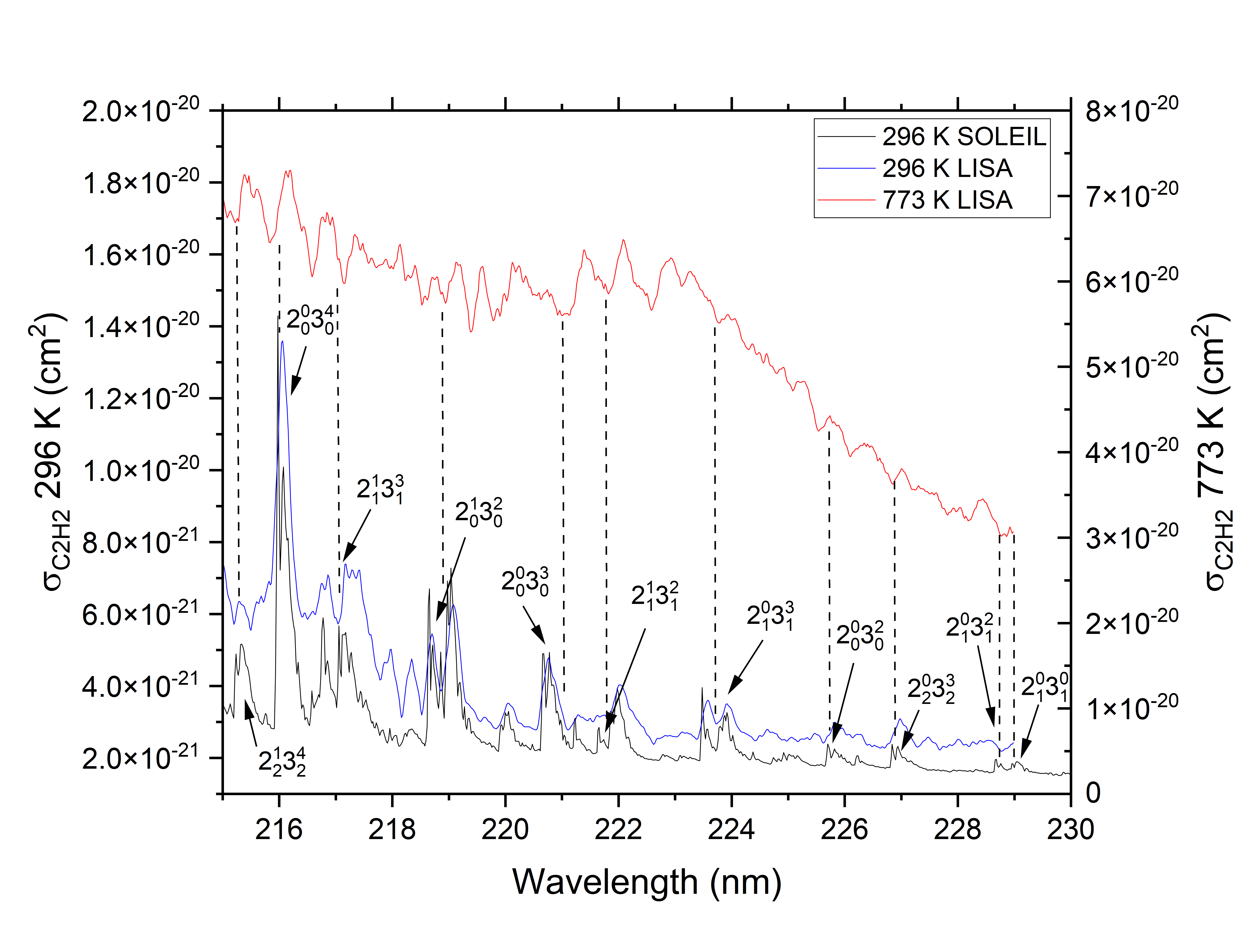}
    \caption {Long wavelength region ($\lambda$ > 215~nm) of the A-X band system of C$_{2}$H$_{2}$ measured at ambient temperature at SOLEIL (black) and LISA (blue) as well as measured at 773~K at LISA (red). Absorption bands in the $\nu_3$ and $\nu_2$ + $\nu_3$ vibrational modes are identified based on the studies of \citet{Watsonetal.1982, VanCraenetal.1985}.}
    \label{Figure_7}
\end{figure}

This more important increase in the absorption cross section beyond 195~nm agrees with the presence of numerous hot bands identified in the A $\leftarrow$ X bands system by \citet{Watsonetal.1982, VanCraenetal.1985} and that we also observed in our spectra. Fig. \ref{Figure_7} presents the identification of numerous absorption bands of the A-X band system between 215 and 230~nm in our spectra measured at LISA and SOLEIL at ambient temperature. These bands involved the $\nu_4$ vibrational mode in the X$^1\Sigma_g^+$ ground state and the $\nu_3$ and $\nu_2$ + $\nu_3$ vibrational modes in the A$^1$A$_u$ first excited state. We observed the presence of hot bands originating from the vibrational levels one and two of the $\nu_4$ vibrational mode in the ground state. Additional hot bands were identified in the studies of \citet{Watsonetal.1982, VanCraenetal.1985} but their intensities are very weak and cannot be observe in our spectra. As is shown in Fig. \ref{Figure_7}, at 773~K, we observed an increase in the continuum intensity compared to ambient temperature and a change in the contrast between cold and hot bands. Indeed, we observe an increase in the intensity of the hot bands regardless of the cold bands. Unfortunately, as the spectra measured by \citet{Zabetietal.2017} are not available, a full and detailed comparison with these data at high temperature is not possible. Nevertheless, we can observe that the cross section at 200~nm measured by these authors at 296~K ($\sim$3 $\times$ 10$^{-20}$ cm$^2$) and 565~K ($\sim$4.5 $\times$ 10$^{-20}$ cm$^2$) are lower by a factor of three compared to those measured at LISA (i.e., at 296~K, $\sim$7.5 $\times$ 10$^{-20}$ cm$^2$ and at 573~K $\sim$1.5 $\times$ 10$^{-19}$ cm$^2$). Finally, we observe that at 773~K, from 215 to 220~nm, the \emph{F$_T$} increases much faster than at lower temperatures, but as was discussed earlier, the thermal decomposition of C$_{2}$H$_{2}$ could bias the results at this higher temperature.

\begin{figure}[h]
    \centering
    \includegraphics [trim=50 0 80 0,clip,width=\hsize] {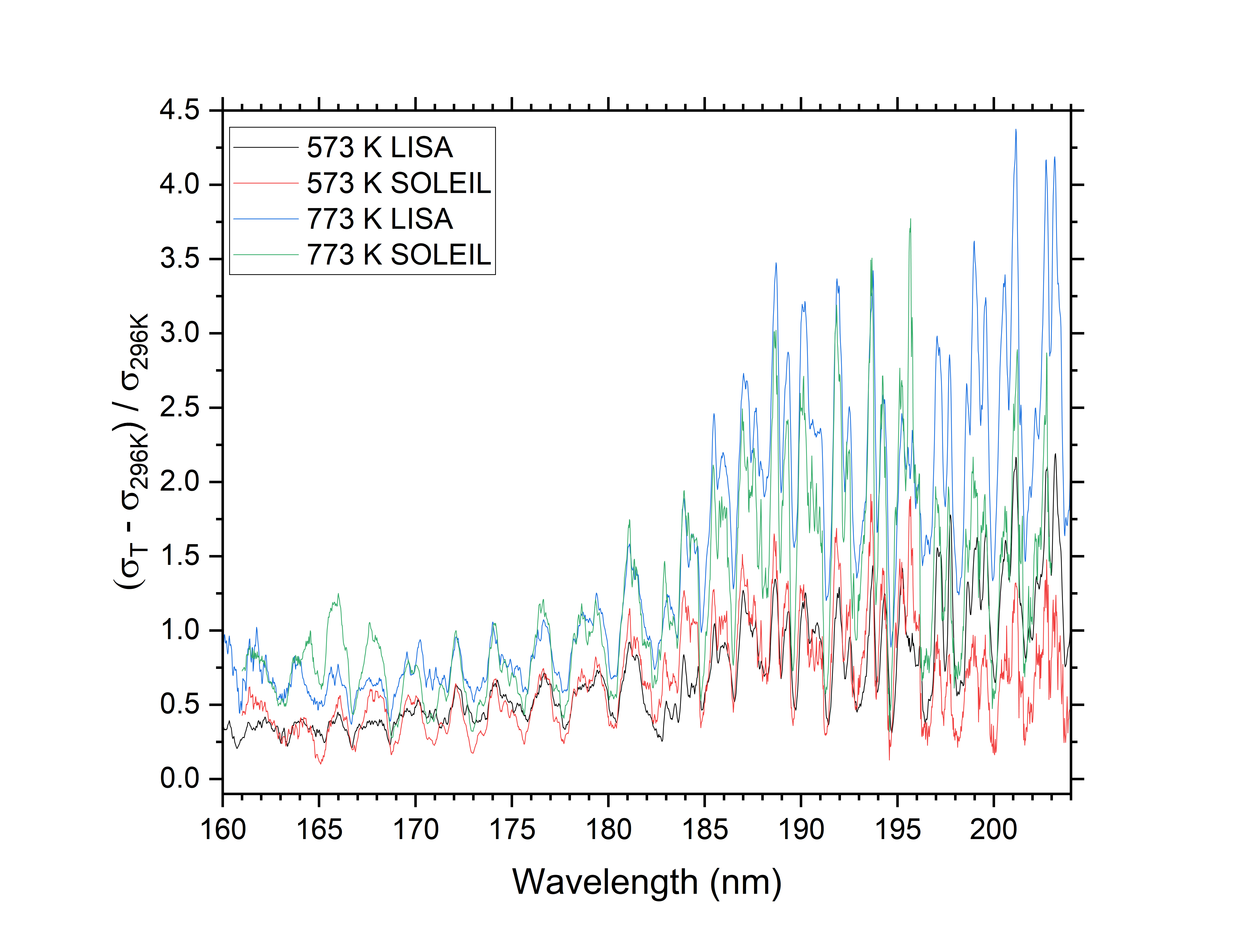}
    \caption {Comparison of the factor of changes of the absorption cross section of C$_{2}$H$_{2}$ for the measurements made at 573 and 773~K at LISA and at SOLEIL from 160 to 204~nm.}
    \label{Figure_8}
\end{figure}

Figure \ref{Figure_8} presents a comparison of \emph{F$_T$} values calculated at 573 and 773~K from 160 to 204~nm using LISA and SOLEIL data. At 573~K, \emph{F$_T$} values from LISA data agree well with those from SOLEIL data. At 773~K, the \emph{F$_T$} values calculated from the two datasets agree up to 195~nm. However, for longer wavelengths, we observe some discrepancies, where \emph{F$_T$} being more significant in LISA data than in SOLEIL ones. These discrepancies can be explained by the variation in the absorption cross section measured at ambient temperature (see Sect. \ref{subsubsec:New data obtained at LISA and SOLEIL facilities}), which is inferred in the calculation of \emph{F$_T$}. 

\subsection{Application to exoplanet atmospheres } \label{subsec:Application to exoplanet atmospheres}
\subsubsection{1D thermo-photochemical model} \label{subsubsec:1D thermo-photochemical model}

We used the 1D thermo-photochemical model FRECKLL \citep{Al-Refaieetal.2024, Veilletetal.2024} to study and quantify the impact of the new C$_{2}$H$_{2}$ absorption cross sections data on the predicted atmospheric composition of exoplanets. We chose to model a hypothetical hot-Jupiter with a solar metallicity \citep{Lodders.2010} and a C/O ratio twice the solar one. From a pure modeling point of view, such an assumption allows us to obtain a high amount of acetylene in the atmosphere of the planet, which is essential to evaluate the effect of our new data. Indeed, several studies have shown that the C$_{2}$H$_{2}$ abundance is enhanced in warm exoplanet atmospheres for a C/O ratio greater than one \citep{Madhusudhan.2012, Mosesetal.2013b, Venotetal.2015, Rocchettoetal.2016}. Such an assumption is also relevant given the large variety of elemental chemical abundances possible for exoplanet atmospheres \citep{Madhusudhanetal.2011, Mosesetal.2013a, Huetal.2015, Cridlandetal.2019, Tsaietal.2021, GuzmanMesaetal.2022}.

We used the thermal profile published by \citet{Venotetal.2015} and calculated with the analytical model of \citet{Parmentieretal.2014}, with an irradiation temperature (as defined by these authors) of 2303~K, leading to a high-altitude atmospheric temperature of 1500~K (Figure 2 of \citealt{Venotetal.2015}). While this temperature is higher than the one at which the cross section of C$_{2}$H$_{2}$ was measured, we made this choice to ensure an important amount of C$_{2}$H$_{2}$ in the upper atmosphere and thus optimize our analysis. To this end, we also tried to maximize the photolysis process and considered that our planet was orbiting a F-type star (Sect. \ref{subsubsec:Stellar irradiation}). Our case-study planet has a radius R = 1.25 R$_{\rm{Jup}}$, a mass m = 1.5 M$_{\rm{Jup}}$, and a gravity g = 25~m~s$^{-2}$. Given the stellar irradiation, the semi-major axis has been fixed to a = 0.05735~AU to obtain an irradiation temperature of 2303~K. We assumed an Eddy diffusion coefficient constant with altitude Kzz = 10$^8$~cm$^2$~s$^{-1}$. We detailed hereafter the chemical scheme, and the irradiation spectra used in our model.

 \subsubsection{Chemical scheme} \label{subsubsec:Chemical scheme}

The model implements the chemical scheme from \citet{Veilletetal.2024}, which includes 175 species composed of C, N, H, and O elements and involved in 2550 reactions (1237 reversible and 76 irreversible). The scheme also includes 65 reactions of photodissociation. This scheme has been validated for a large range of pressures and temperatures through an extensive comparison with experimental combustion data \citep{Veilletetal.2024}. It is therefore very robust and well suited to model warm atmospheres.

\subsubsection{Stellar irradiation} \label{subsubsec:Stellar irradiation}

The stellar spectrum of the exoplanet host star is an important input that allows one to model photolysis process, and thus calculate the photodissociation rates of atmospheric molecules. As in \citet{venotetal2018}, we used a stellar spectrum of HD 128167 (F2V), which was constructed thanks to observational data of this star from 115 to 900~nm \citep{Seguraetal.2003}. For the shorter wavelengths (1-114~nm), we used data from the Sun \citep{Thuillieretal.2004}, scaled to the effective temperature and radius of HD 128167 (T = 6723~K and R = 1.434~R$_{\rm{Sun}}$).

 \subsubsection{C$_{2}$H$_{2}$ absorption cross section} \label{subsubsec:C2H2 absorption cross section}

We modeled the chemical composition of the atmosphere of our hypothetical planet using the absorption cross section of C$_{2}$H$_{2}$ measured in this study (see Sect. \ref{subsec:Evolution of C2H2 absorption cross section with the temperature}). To evaluate the effect of the thermal dependence of absorption cross sections, we ran two sets of simulations. From 7 to 115~nm, we used in both cases the absorption cross section of C$_{2}$H$_{2}$ measured at ambient temperature by \citet{Cooperetal.1995}. In the first simulation, we used the C$_{2}$H$_{2}$ absorption cross sections obtained at LISA at ambient temperature for the spectral range of 115–228~nm, whereas in the second simulation, we employed high temperature absorption cross sections measured at 773~K. For wavelengths longer than 228~nm, we assumed zero absorption for C$_{2}$H$_{2}$. We did not change the absorption cross sections of the other molecules between the simulations. To reduce the computational time, the model used a resolution of 1~nm for both the stellar flux and the absorption cross sections. Our data have thus been reduced to this resolution. At longer wavelengths, \citet{Benilanetal.2000} studied the dependence of the photo-dissociation rate of  C$_{2}$H$_{2}$ on the spectral resolution of its absorption cross sections within the 185–235~nm spectral range. They demonstrated that, while the calculated photo-dissociation rate of  C$_{2}$H$_{2}$ varies with the spectral resolution, this variation is very minor, within less than 1\%. Accordingly, the high resolution does not introduce significant uncertainties into the resulting values calculated by the model.

\subsubsection{Photodissociation rate of C$_{2}$H$_{2}$} \label{subsubsec:Photodissociation rate of C2H2}

To quantify the effect of the change of the absorption cross section with the temperature on the photo-dissociation and subsequently on the modeled molecular abundances, we can compare the photodissociation rate (s$^{-1}$) of C$_{2}$H$_{2}$ in our two different simulations described above. In our model, following the assumption of \citet{Hebrardetal.2013, Heaysetal.2017, Vuittonetal.2019} based on the quantum yield measurements of \citet{Lauteretal.2002}, we consider only one C$_{2}$H$_{2}$ photodissociation route:

\begin{equation}
\begin{split}
\rm C_{2}H_{2} + h\nu & \longrightarrow \rm C_{2}H + H,
    \label{Eq.3}
\end{split}
\end{equation}
with a quantum yield of one from 120 to 217 nm. From 72 to 120 nm, we used the quantum yield recommended by \citet{HuebnerandMukherjee.2015}.

\begin{figure}[h]
    \centering
    \includegraphics [width=\hsize] {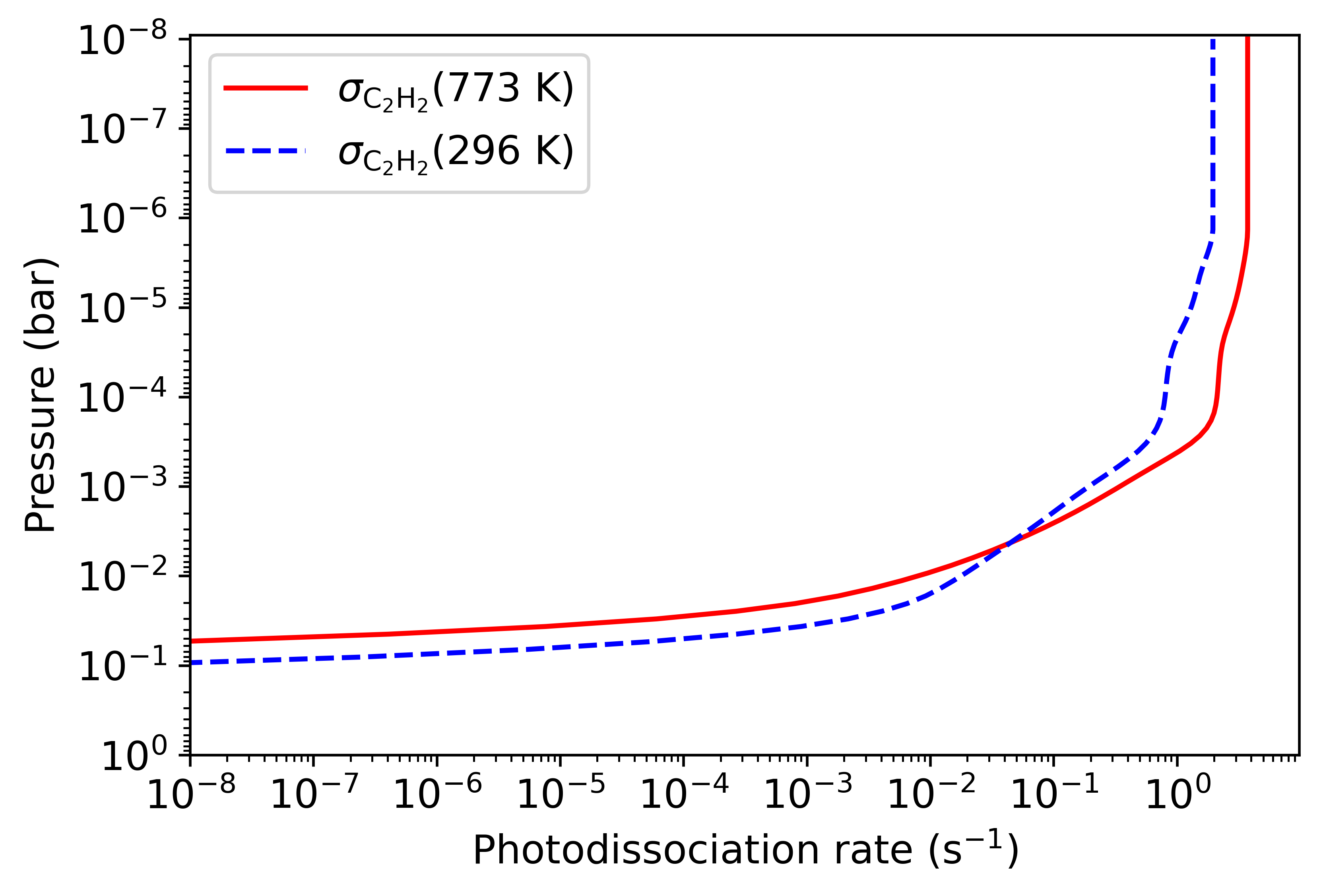}
    \caption {Photodissociation rates of C$_{2}$H$_{2}$ in our atmospheric model using the absorption cross section of C$_{2}$H$_{2}$ measured at 296~K (dashed blue line) and 773~K (full red line).}
    \label{Figure_9}
\end{figure}

\begin{figure}[h]
    \centering
    \includegraphics [width=\hsize] {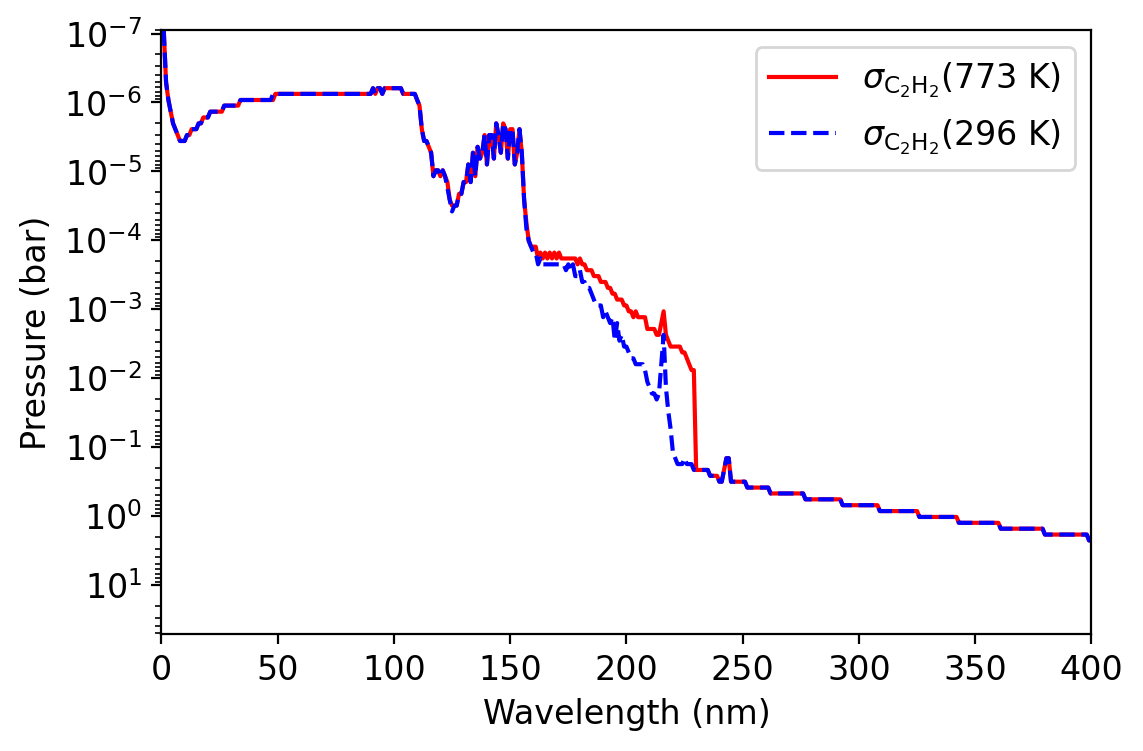}
    \caption {Penetration of the actinic flux in the atmosphere (level where the optical depth is equal to 1) when using $\sigma_{\rm{C_2H_2}}$ (296~K) (dashed blue line) and $\sigma_{\rm{C_2H_2}}$ (773~K) (full red line). This takes into account absorption and diffusion by the different chemical species presents in the atmosphere.}
    \label{Figure_10}
\end{figure}

Figure \ref{Figure_9} presents the photodissociation rate of C$_{2}$H$_{2}$ as a function of pressure in our two simulations. For pressures lower than 10$^{-2}$ bar, we emphasize that the photodissociation rate of C$_{2}$H$_{2}$ is three times higher when using the absorption cross section at 773~K. This increase in the photodissociation rate in the upper atmosphere is attributed to the rise in the absorption cross section with temperature. In addition, we observed that when using $\sigma_{\rm{C_2H_2}}$ (773~K), the photodissociation rate becomes negligible below P $\sim$ 5 × 10$^{-1}$ bar, indicating that no photodissociation of C$_{2}$H$_{2}$ occurs at deeper atmospheric levels. However, when using $\sigma_{\rm{C_2H_2}}$ (296~K), C$_{2}$H$_{2}$ photodissociation occurs down to $\sim$ 2 × 10$^{-1}$ bar. This can be explained by a change in the penetration of the actinic flux into the atmosphere, depending on the absorption cross section of acetylene used. Figure \ref{Figure_10} presents, for our two simulations, the altitude at which the optical depth of atmosphere reaches a value of one as a function of wavelength, considering the absorption and diffusion processes by various chemical species present in the atmosphere. In the spectral range of 150-228~nm, where the absorption is primarily dominated by C$_{2}$H$_{2}$, we observe that the actinic flux penetrates less deeply in the atmosphere when using data at 773~K. This is consequently due to the increases in the absorption cross section of C$_{2}$H$_{2}$ at higher temperature. It should be noted that the discontinuity observed at 228~nm in the simulation using $\sigma_{\rm{C_2H_2}}$ (773~K) arises from assumption of zero absorption by acetylene at longer wavelengths. Therefore, the opacity of the atmosphere calculated by the model passes abruptly from a domain governed by C$_{2}$H$_{2}$ absorption to one potentially dominated by Rayleigh diffusion. Even if not observed in this study, such an impact on the penetration of the stellar flux could have significant consequences for the abundance of other absorbing species, resulting in lower degree of  photodissociation, and thus higher abundances. For instance, the self-shielding mechanism could explain isotopic ratios anomalies in planetary atmospheres due to the isotope fractionation \citep{Chakrabortyetal.2014, Lyons2020, Lyonsetal.2005}. While we did not specifically attempt to investigate this mechanism of shielding, such an effect was observed with CO$_2$ in a similar study \citep{venotetal2018} when analyzing chemical pathways.

\subsubsection{Composition at the steady state} \label{subsubsec:Composition at the steady state}

\begin{figure}[h]
    \centering
    \includegraphics [width=\hsize] {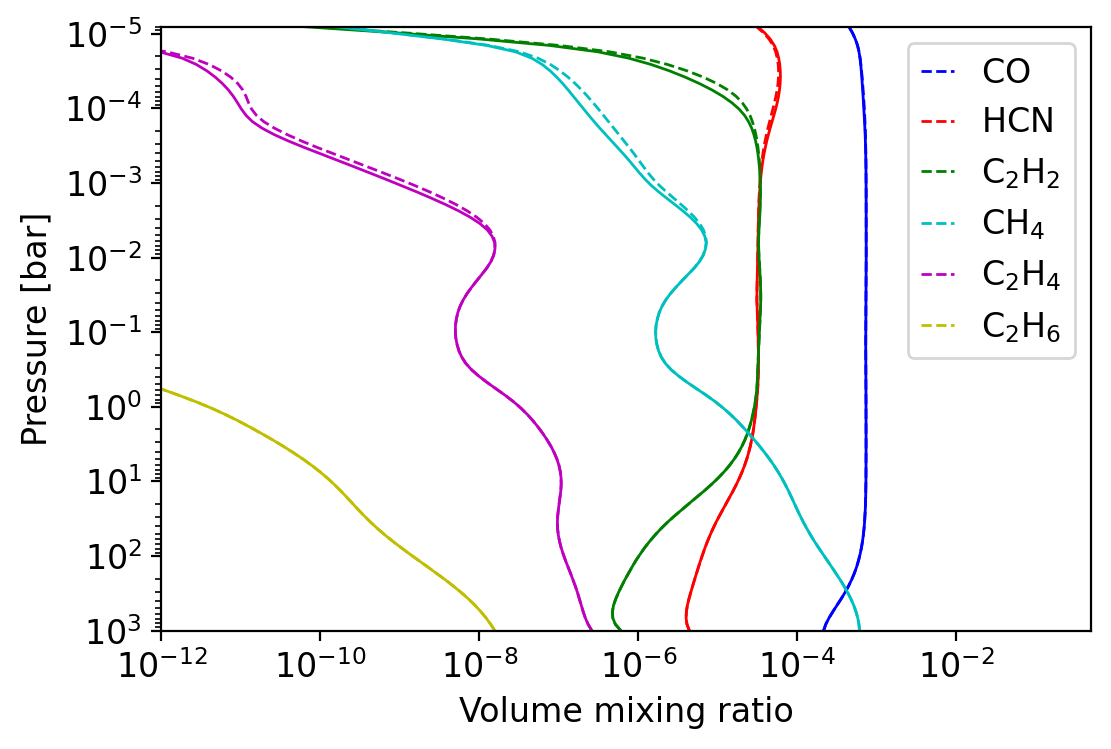}
    \caption {Abundance profiles for the main hydrocarbons at steady state in the atmosphere of our hypothetical planet. Solid lines correspond to the abundances calculated using the absorption cross section of C$_{2}$H$_{2}$ measured at 773~K, and dotted lines correspond to 296~K.}
    \label{Figure_11}
\end{figure}

Figure \ref{Figure_11} presents the vertical volume mixing ratio profiles for the main hydrocarbons, as well as CO, at steady state for the two simulations. After dihydrogen (H$_2$) and helium (He), which are not shown in the figure, the atmosphere is dominated by CO ($\sim$10$^{-3}$), hydrogen cyanide (HCN), C$_{2}$H$_{2}$, and CH$_4$. This is a consequence of the high C/O ratio chosen for our simulations. A solar C/O ratio would have led to an amount of H$_2$O almost similar to that of CO, and a very low abundance of hydrocarbons \citep{Venotetal.2015}. In our simulations, the other C$_2$ hydrocarbons, that is ethylene (C$_{2}$H$_{4}$) and ethane (C$_{2}$H$_{6}$), are less important and have volume mixing ratios lower than 10$^{-6}$.

We observe in Fig. \ref{Figure_11} that the abundance profile of C$_{2}$H$_{2}$ is slightly modified in the high atmosphere when the absorption cross section measured at 773~K is used instead of the one at 296~K, with a decrease of 40\% around 5 $\times$ 10$^{-5}$ bar. This finding is similar to that obtained for CO$_2$ by  \citet{venotetal2018}. Indeed, they observed that the change in CO$_2$ abundance was not significant even though the absorption cross section of CO$_2$ increases by four orders of magnitude in the longer wavelength spectral domain at 773~K compared to 296~K. In Fig. \ref{Figure_11}, we note also that other species undergo variations in abundance, such as HCN, which becomes slightly more abundant when using $\sigma_{\rm{C_2H_2}}$ at 773~K and most notably CH$_{4}$ and C$_{2}$H$_{4}$, where their abundances diminished down to 10$^{-2}$~bar, showing the strong coupling between molecules through chemical reactions.

\section{Conclusions} \label{sec:Conclusions}

We have measured experimentally for the first time the thermal dependence of the VUV absorption cross section of C$_{2}$H$_{2}$ at seven temperatures ranging from 296 to 793~K and over a large spectral range extending from 115 to 230~nm.

We found that the absolute absorption cross section of C$_{2}$H$_{2}$ increases with temperature. We observed an increase in the intensity of the continuum and a decrease in the intensity of the absorption bands as the gas temperature increase, which result in an overall increase in the absolute absorption cross section of C$_{2}$H$_{2}$ with temperature. From 116 to 185~nm, the continuum increases is relatively constant across the whole wavelengths range at a given temperature (e.g., a \emph{F$_T$} of 0.5 at 773~K), while the variations in certain bands can be more important (up to a \emph{F$_T$} of 3.5 at 773~K). Beyond 185~nm, the increase is more important and reaches a maximum of a factor of 20 around 220~nm, which agrees with the presence of numerous hot bands in this region. In addition to these general variations, we observe at 473~K (and at higher temperatures) a new absorption band at $\sim$156.7~nm, whose intensity increases together with the temperature. We identify it as a hot band, which originates from the vibrational mode $\nu_2$ ($\nu_2$ = 1973.8 cm$^{-1}$).

To go further, we quantified the impact of these new data on the prediction of exoplanet atmospheres composition using the 1D thermo-photochemical model FRECKLL. After modeling a hypothetical exoplanet atmosphere, we found that the abundance profile of C$_{2}$H$_{2}$ is slightly changed when the absorption cross section of C$_{2}$H$_{2}$ measured at 773~K is used instead of the one at 296~K, with a decrease of 40\% around 5 $\times$ 10$^{-5}$ bar. In addition, these changes in the absorption cross sections of C$_{2}$H$_{2}$ affect the penetration of the actinic flux through the atmosphere from 150 to 230~nm, resulting in an attenuation of the flux at higher altitudes when using the cross section measured at 773~K compared to the ones measured at 296~K. Although we did not compute it, we do not expect this slight change of the C$_{2}$H$_{2}$ abundance profile to have a visible impact on the synthetic spectrum of the simulated atmosphere, and to thus be visible in observational data.

Finally, the present study only investigates the effect of the thermal dependence of the absorption cross section of C$_{2}$H$_{2}$ on simulated atmospheric compositions. However, similar changes in the absorption cross section with temperature have been observed for CO$_2$ \citep{Venotetal.2013, venotetal2018} and can be expected for numerous species present in exoplanet atmospheres. It is therefore essential to measure the thermal dependence of the VUV absorption cross section of all major species observed in exoplanet atmospheres, knowing that the effect of photochemistry on the composition of an exoplanet atmosphere has now been observed with JWST \citep{Tsaietal.2023, Dyreketal.2024}.

\begin{acknowledgements}
    This work is supported by the ANR project ‘EXACT’ (ANR-21-CE49-0008-01), the Centre National d'Etudes Spatiales (CNES), and the CNRS/INSU Programme National de Planétologie (PNP). B.F. thanks the Université Paris-Est Créteil (UPEC) for funding support (postdoctoral grant).
\end{acknowledgements}

\bibliographystyle{aa} 
\bibliography{aa51638} 

\begin{thebibliography}{91}
\expandafter\ifx\csname natexlab\endcsname\relax\def\natexlab#1{#1}\fi

\bibitem[{Aghsaee {et~al.}(2014)Aghsaee, Dürrstein, Herzler, Böhm, Fikri, \& Schulz}]{Aghsaeeetal.2014}
Aghsaee, M., Dürrstein, S.~H., Herzler, J., {et~al.} 2014, Combustion and Flame, 161, 2263

\bibitem[{Al-Refaie {et~al.}(2024)Al-Refaie, Venot, Changeat, \& Edwards}]{Al-Refaieetal.2024}
Al-Refaie, A.~F., Venot, O., Changeat, Q., \& Edwards, B. 2024, The Astrophysical Journal, 967, 132

\bibitem[{Alderson {et~al.}(2023)Alderson, Wakeford, Alam, Batalha, Lothringer, Adams~Redai, Barat, Brande, Damiano, Daylan, Espinoza, Flagg, Goyal, Grant, Hu, Inglis, Lee, Mikal-Evans, Ramos-Rosado, Roy, Wallack, Batalha, Bean, Benneke, Berta-Thompson, Carter, Changeat, Colón, Crossfield, Désert, Foreman-Mackey, Gibson, Kreidberg, Line, López-Morales, Molaverdikhani, Moran, Morello, Moses, Mukherjee, Schlawin, Sing, Stevenson, Taylor, Aggarwal, Ahrer, Allen, Barstow, Bell, Blecic, Casewell, Chubb, Crouzet, Cubillos, Decin, Feinstein, Fortney, Harrington, Heng, Iro, Kempton, Kirk, Knutson, Krick, Leconte, Lendl, MacDonald, Mancini, Mansfield, May, Mayne, Miguel, Nikolov, Ohno, Palle, Parmentier, Petit dit de~la Roche, Piaulet, Powell, Rackham, Redfield, Rogers, Rustamkulov, Tan, Tremblin, Tsai, Turner, de~Val-Borro, Venot, Welbanks, Wheatley, \& Zhang}]{Aldersonetal.2023}
Alderson, L., Wakeford, H.~R., Alam, M.~K., {et~al.} 2023, Nature, 614, 664

\bibitem[{Baeyens {et~al.}(2022)Baeyens, Konings, Venot, Carone, \& Decin}]{Baeyensetal.2022}
Baeyens, R., Konings, T., Venot, O., Carone, L., \& Decin, L. 2022, Monthly Notices of the Royal Astronomical Society, 512, 4877

\bibitem[{Benson(1992)}]{Benson1992}
Benson, S.~W. 1992, International Journal of Chemical Kinetics, 24, 217

\bibitem[{Boyé {et~al.}(2004)Boyé, Campos, Fillion, Douin, Shafizadeh, \& Gauyacq}]{Boyeetal.2004}
Boyé, S., Campos, A., Fillion, J.-H., {et~al.} 2004, Comptes Rendus Physique, 5, 239

\bibitem[{Bénilan {et~al.}(1995)Bénilan, Andrieux, \& Bruston}]{Benilanetal.1995}
Bénilan, Y., Andrieux, D., \& Bruston, P. 1995, Geophysical Research Letters, 22, 897

\bibitem[{Bénilan {et~al.}(2000)Bénilan, Smith, Jolly, \& Raulin}]{Benilanetal.2000}
Bénilan, Y., Smith, N., Jolly, A., \& Raulin, F. 2000, Planetary and Space Science, 48, 463

\bibitem[{Chakraborty {et~al.}(2014)Chakraborty, Muskatel, Jackson, Ahmed, Levine, \& Thiemens}]{Chakrabortyetal.2014}
Chakraborty, S., Muskatel, B.~H., Jackson, T.~L., {et~al.} 2014, Proc Natl Acad Sci USA, 111, 14704

\bibitem[{Chen {et~al.}(1991)Chen, Judge, Robert~Wu, Caldwell, White, \& Wagener}]{Chenetal.1991}
Chen, F., Judge, D.~L., Robert~Wu, C.~Y., {et~al.} 1991, Journal of Geophysical Research: Planets, 96, 17519

\bibitem[{Chen \& Wu(2004)}]{ChenandWu.2004}
Chen, F.~Z. \& Wu, C. Y.~R. 2004, Journal of Quantitative Spectroscopy and Radiative Transfer, 85, 195

\bibitem[{Cheng {et~al.}(2011)Cheng, Chen, Lu, Chen, Alam, Chou, \& Lin}]{Chengetal.2011}
Cheng, B.-M., Chen, H.-F., Lu, H.-C., {et~al.} 2011, The Astrophysical Journal Supplement Series, 196, 3

\bibitem[{Cheng {et~al.}(2006)Cheng, Lu, Chen, Bahou, Lee, Mebel, Lee, Liang, \& Yung}]{Chengetal.2006}
Cheng, B.-M., Lu, H.-C., Chen, H.-K., {et~al.} 2006, The Astrophysical Journal, 647, 1535

\bibitem[{Chubb {et~al.}(2024)Chubb, Robert, Sousa-Silva, Yurchenko, Allard, Boudon, Buldyreva, Bultel, Coustenis, Foltynowicz, Gordon, Hargreaves, Helling, Hill, Hrodmarsson, Karman, Lecoq-Molinos, Migliorini, Rey, Richard, Sadiek, Schmidt, Sokolov, Stefani, Tennyson, Venot, Wright, Arenales-Lope, Barstow, Bocchieri, Carrasco, Dubey, Egorov, Muñoz, Gharib-Nezhad, Gkouvelis, Grübel, Irwin, Knížek, Lewis, Lodge, Ma, Martins, Molaverdikhani, Morello, Nikitin, Panek, Rengel, Rinaldi, Skinner, Tinetti, van Kempen, Yang, \& Zingales}]{Chubbetal.2024}
Chubb, K.~L., Robert, S., Sousa-Silva, C., {et~al.} 2024, RAS Techniques and Instruments, rzae039

\bibitem[{Colket {et~al.}(1989)Colket, Seery, \& Palmer}]{Colketetal.1989}
Colket, M.~B., Seery, D.~J., \& Palmer, H.~B. 1989, Combustion and Flame, 75, 343

\bibitem[{Cooper {et~al.}(1995)Cooper, Burton, \& Brion}]{Cooperetal.1995}
Cooper, G., Burton, G.~R., \& Brion, C.~E. 1995, Journal of Electron Spectroscopy and Related Phenomena, 73, 139

\bibitem[{Cridland {et~al.}(2019)Cridland, van Dishoeck, Alessi, \& Pudritz}]{Cridlandetal.2019}
Cridland, A.~J., van Dishoeck, E.~F., Alessi, M., \& Pudritz, R.~E. 2019, A\&A, 632, A63

\bibitem[{de~Oliveira {et~al.}(2011)de~Oliveira, Roudjane, Joyeux, Phalippou, Rodier, \& Nahon}]{deOliveiraetal.2011}
de~Oliveira, N., Roudjane, M., Joyeux, D., {et~al.} 2011, Nature Photonics, 5, 149

\bibitem[{Drummond {et~al.}(2019)Drummond, Carter, Hébrard, Mayne, Sing, Evans, \& Goyal}]{Drummondetal.2019}
Drummond, B., Carter, A.~L., Hébrard, E., {et~al.} 2019, Monthly Notices of the Royal Astronomical Society, 486, 1123

\bibitem[{Duran {et~al.}(1988)Duran, Amorebieta, \& Colussi}]{Duranetal.1988}
Duran, R.~P., Amorebieta, V.~T., \& Colussi, A.~J. 1988, The Journal of Physical Chemistry, 92, 636, doi: 10.1021/j100314a014

\bibitem[{Dyrek {et~al.}(2024)Dyrek, Min, Decin, Bouwman, Crouzet, Mollière, Lagage, Konings, Tremblin, Güdel, Pye, Waters, Henning, Vandenbussche, Ardevol~Martinez, Argyriou, Ducrot, Heinke, van Looveren, Absil, Barrado, Baudoz, Boccaletti, Cossou, Coulais, Edwards, Gastaud, Glasse, Glauser, Greene, Kendrew, Krause, Lahuis, Mueller, Olofsson, Patapis, Rouan, Royer, Scheithauer, Waldmann, Whiteford, Colina, van Dishoeck, Östlin, Ray, \& Wright}]{Dyreketal.2024}
Dyrek, A., Min, M., Decin, L., {et~al.} 2024, Nature, 625, 51

\bibitem[{Es-sebbar \& Farooq(2014)}]{ESsebbar2014}
Es-sebbar, E. \& Farooq, A. 2014, Journal of Quantitative Spectroscopy and Radiative Transfer, 149, 241

\bibitem[{Foo \& Innes(1973)}]{FooandInes.1973}
Foo, P.~D. \& Innes, K.~K. 1973, Chemical Physics Letters, 22, 439

\bibitem[{Fortney {et~al.}(2019)Fortney, Robinson, Domagal-Goldman, Genio, Gordon, Gharib-Nezhad, Lewis, Sousa-Silva, Airapetian, Drouin, Hargreaves, Huang, Karman, Ramirez, Rieker, Tennyson, Wordsworth, Yurchenko, Johnson, Lee, Marley, Dong, Kane, López-Morales, Fauchez, Lee, Sung, Haghighipour, Horst, Gao, Kao, Dressing, Lupu, Savin, Fleury, Venot, Ascenzi, Milam, Linnartz, Gudipati, Gronoff, Salama, Gavilan, Bouwman, Turbet, Benilan, Henderson, Batalha, Jensen-Clem, Lyons, Freedman, Schwieterman, Goyal, Mancini, Irwin, Desert, Molaverdikhani, Gizis, Taylor, Lothringer, Pierrehumbert, Zellem, Batalha, Rugheimer, Lustig-Yaeger, Hu, Kempton, Arney, Line, Alam, Moses, Iro, Kreidberg, Blecic, Louden, Mollière, Stevenson, Swain, Bott, Madhusudhan, Krissansen-Totton, Deming, Kitiashvili, Shkolnik, Rustamkulov, Rogers, \& Close}]{Fortneyetal.2019}
Fortney, J., Robinson, T.~D., Domagal-Goldman, S., {et~al.} 2019, Astro2020: Decadal Survey on Astronomy and Astrophysics, 2020, 146

\bibitem[{Gedanken \& Schnepp(1976)}]{GedankenandSchnepp.1976}
Gedanken, A. \& Schnepp, O. 1976, Chemical Physics Letters, 37, 373

\bibitem[{Giacobbe {et~al.}(2021)Giacobbe, Brogi, Gandhi, Cubillos, Bonomo, Sozzetti, Fossati, Guilluy, Carleo, Rainer, Harutyunyan, Borsa, Pino, Nascimbeni, Benatti, Biazzo, Bignamini, Chubb, Claudi, Cosentino, Covino, Damasso, Desidera, Fiorenzano, Ghedina, Lanza, Leto, Maggio, Malavolta, Maldonado, Micela, Molinari, Pagano, Pedani, Piotto, Poretti, Scandariato, Yurchenko, Fantinel, Galli, Lodi, Sanna, \& Tozzi}]{Giacobbetal.2021}
Giacobbe, P., Brogi, M., Gandhi, S., {et~al.} 2021, Nature, 592, 205

\bibitem[{Grosch {et~al.}(2015)Grosch, Fateev, \& Clausen}]{Groschetal.2015}
Grosch, H., Fateev, A., \& Clausen, S. 2015, Journal of Quantitative Spectroscopy and Radiative Transfer, 154, 28

\bibitem[{Guzmán-Mesa {et~al.}(2022)Guzmán-Mesa, Kitzmann, Mordasini, \& Heng}]{GuzmanMesaetal.2022}
Guzmán-Mesa, A., Kitzmann, D., Mordasini, C., \& Heng, K. 2022, Monthly Notices of the Royal Astronomical Society, 513, 4015

\bibitem[{Heays {et~al.}(2017)Heays, Bosman, \& van Dishoeck}]{Heaysetal.2017}
Heays, A.~N., Bosman, A.~D., \& van Dishoeck, E.~F. 2017, A\&A, 602, A105

\bibitem[{Herbert {et~al.}(1987)Herbert, Sandel, Yelle, Holberg, Broadfoot, Shemansky, Atreya, \& Romani}]{Herbertetal1987}
Herbert, F., Sandel, B.~R., Yelle, R.~V., {et~al.} 1987, Journal of Geophysical Research: Space Physics, 92, 15093

\bibitem[{Herzberg(1966)}]{Herzberg.1966}
Herzberg, G. 1966, Molecular spectra and molecular structure. Vol.3: Electronic spectra and electronic structure of polyatomic molecules

\bibitem[{Hu {et~al.}(2015)Hu, Seager, \& Yung}]{Huetal.2015}
Hu, R., Seager, S., \& Yung, Y.~L. 2015, The Astrophysical Journal, 807, 8

\bibitem[{Huebner \& Mukherjee(2015)}]{HuebnerandMukherjee.2015}
Huebner, W.~F. \& Mukherjee, J. 2015, Planetary and Space Science, 106, 11

\bibitem[{Hébrard {et~al.}(2013)Hébrard, Dobrijevic, Loison, Bergeat, Hickson, \& Caralp}]{Hebrardetal.2013}
Hébrard, E., Dobrijevic, M., Loison, J.~C., {et~al.} 2013, A\&A, 552, A132

\bibitem[{Hörst(2017)}]{Horst2017}
Hörst, S.~M. 2017, Journal of Geophysical Research: Planets, 122, 432

\bibitem[{Kawashima \& Masahiro(2018)}]{KawashimaandMasahiro.2018}
Kawashima, Y. \& Masahiro, I. 2018, The Astrophysical Journal, 853, 7

\bibitem[{Kern {et~al.}(1991)Kern, Xie, Chen, \& Kiefer}]{Kernetal.1991}
Kern, R.~D., Xie, K., Chen, H., \& Kiefer, J.~H. 1991, Symposium (International) on Combustion, 23, 69

\bibitem[{Line {et~al.}(2011)Line, Gautam, Pin, D., \& Yuk~L.}]{Lineteal.2011}
Line, M.~R., Gautam, V., Pin, C., D., A., \& Yuk~L., Y. 2011, The Astrophysical Journal, 738, 32

\bibitem[{Line {et~al.}(2010)Line, Liang, \& Yung}]{Lineetal.2010}
Line, M.~R., Liang, M.~C., \& Yung, Y.~L. 2010, The Astrophysical Journal, 717, 496

\bibitem[{Liu {et~al.}(2021)Liu, Chu, Jocher, Smith, Lengyel, \& Green}]{Liuetal.2021}
Liu, M., Chu, T.-C., Jocher, A., {et~al.} 2021, International Journal of Chemical Kinetics, 53, 27

\bibitem[{Lodders(2010)}]{Lodders.2010}
Lodders, K. 2010, in Principles and Perspectives in Cosmochemistry, ed. A.~Goswami \& B.~E. Reddy (Springer Berlin Heidelberg), 379--417

\bibitem[{Lyons(2020)}]{Lyons2020}
Lyons, J. 2020, Geochimica et Cosmochimica Acta, 282, 177

\bibitem[{Lyons \& Young(2005)}]{Lyonsetal.2005}
Lyons, J.~R. \& Young, E.~D. 2005, Nature, 435, 317

\bibitem[{Läuter {et~al.}(2002)Läuter, Lee, Jung, Vatsa, Mittal, \& Volpp}]{Lauteretal.2002}
Läuter, A., Lee, K.~S., Jung, K.~H., {et~al.} 2002, Chemical Physics Letters, 358, 314

\bibitem[{Macy(1980)}]{Macy1980}
Macy, W. 1980, Icarus, 41, 153

\bibitem[{Madhusudhan(2012)}]{Madhusudhan.2012}
Madhusudhan, N. 2012, The Astrophysical Journal, 758, 36

\bibitem[{Madhusudhan {et~al.}(2011)Madhusudhan, Mousis, Johnson, \& Lunine}]{Madhusudhanetal.2011}
Madhusudhan, N., Mousis, O., Johnson, T.~V., \& Lunine, J.~I. 2011, The Astrophysical Journal, 743, 191

\bibitem[{Meyn(2000)}]{Meyn}
Meyn, L. 2000, An uncertainty propagation methodology that simplifies uncertainty analyses (American Institute of Aeronautics \& Astronautics)

\bibitem[{{Moos} \& {Clarke}(1979)}]{Moos1979}
{Moos}, H.~W. \& {Clarke}, J.~T. 1979, The Astrophysical Journal Letters, 229, L107

\bibitem[{Moses(2014)}]{Moses.2014}
Moses, J.~I. 2014, Philos Trans A Math Phys Eng Sci, 372, 20130073

\bibitem[{Moses {et~al.}(2013{\natexlab{a}})Moses, Line, Visscher, Richardson, Nettelmann, Fortney, Barman, Stevenson, \& Madhusudhan}]{Mosesetal.2013a}
Moses, J.~I., Line, M.~R., Visscher, C., {et~al.} 2013{\natexlab{a}}, The Astrophysical Journal, 777, 34

\bibitem[{Moses {et~al.}(2013{\natexlab{b}})Moses, Madhusudhan, Visscher, \& Freedman}]{Mosesetal.2013b}
Moses, J.~I., Madhusudhan, N., Visscher, C., \& Freedman, R.~S. 2013{\natexlab{b}}, The Astrophysical Journal, 763, 25

\bibitem[{Moses {et~al.}(2011)Moses, Visscher, Fortney, Showman, Lewis, Griffith, Klippenstein, Shabram, Friedson, Marley, \& Freedman}]{Mosesetal.2011}
Moses, J.~I., Visscher, C., Fortney, J.~J., {et~al.} 2011, The Astrophysical Journal, 737

\bibitem[{Nahon {et~al.}(2012)Nahon, de~Oliveira, Garcia, Gil, Pilette, Marcouillé, Lagarde, \& Polack}]{Nahonetal.2012}
Nahon, L., de~Oliveira, N., Garcia, G.~A., {et~al.} 2012, Journal of Synchrotron Radiation, 19, 508

\bibitem[{Nakayama \& Watanabe(1964)}]{Nakayama.1964}
Nakayama, T. \& Watanabe, K. 1964, The Journal of Chemical Physics, 40, 558

\bibitem[{Nixon(2024)}]{Nixon2024}
Nixon, C.~A. 2024, ACS Earth and Space Chemistry, 8, 406

\bibitem[{Nobre {et~al.}(2008)Nobre, Fernandes, Ferreira~da Silva, Antunes, Almeida, Kokhan, Hoffmann, Mason, Eden, \& Limão-Vieira}]{Nobreetal.2008}
Nobre, M., Fernandes, A., Ferreira~da Silva, F., {et~al.} 2008, Physical Chemistry Chemical Physics, 10, 550

\bibitem[{Orton {et~al.}(1987)Orton, Aitken, Smith, Roche, Caldwell, \& Snyder}]{Orton1987}
Orton, G.~S., Aitken, D.~K., Smith, C., {et~al.} 1987, Icarus, 70, 1

\bibitem[{Parmentier \& Guillot(2014)}]{Parmentieretal.2014}
Parmentier, V. \& Guillot, T. 2014, A\&A, 562, A133

\bibitem[{Poveda(2023)}]{poveda:tel-04718446}
Poveda, M. 2023, Theses, {Universit{\'e} Paris Est}

\bibitem[{Powell {et~al.}(2024)Powell, Feinstein, Lee, Zhang, Tsai, Taylor, Kirk, Bell, Barstow, Gao, Bean, Blecic, Chubb, Crossfield, Jordan, Kitzmann, Moran, Morello, Moses, Welbanks, Yang, Zhang, Ahrer, Bello-Arufe, Brande, Casewell, Crouzet, Cubillos, Demory, Dyrek, Flagg, Hu, Inglis, Jones, Kreidberg, López-Morales, Lagage, Meier~Valdés, Miguel, Parmentier, Piette, Rackham, Radica, Redfield, Stevenson, Wakeford, Aggarwal, Alam, Batalha, Batalha, Benneke, Berta-Thompson, Brady, Caceres, Carter, Désert, Harrington, Iro, Line, Lothringer, MacDonald, Mancini, Molaverdikhani, Mukherjee, Nixon, Oza, Palle, Rustamkulov, Sing, Steinrueck, Venot, Wheatley, \& Yurchenko}]{Powelletal.2024}
Powell, D., Feinstein, A.~D., Lee, E. K.~H., {et~al.} 2024, Nature, 626, 979

\bibitem[{Ranjan {et~al.}(2020)Ranjan, Schwieterman, Harman, Fateev, Sousa-Silva, Seager, \& Hu}]{Ranjanetal.2020}
Ranjan, S., Schwieterman, E.~W., Harman, C., {et~al.} 2020, The Astrophysical Journal, 896, 148

\bibitem[{Ridgway(1974)}]{ridgwayetal.1974}
Ridgway, S.~T. 1974, The Astrophysical Journal, 187, L41

\bibitem[{Rocchetto {et~al.}(2016)Rocchetto, Waldmann, Venot, Lagage, \& Tinetti}]{Rocchettoetal.2016}
Rocchetto, M., Waldmann, I.~P., Venot, O., Lagage, P.~O., \& Tinetti, G. 2016, The Astrophysical Journal, 833, 120

\bibitem[{Rufus {et~al.}(2009)Rufus, Stark, Thorne, Pickering, Blackwell-Whitehead, Blackie, \& Smith}]{Rufusetal.2009}
Rufus, J., Stark, G., Thorne, A.~P., {et~al.} 2009, Journal of Geophysical Research: Planets, 114

\bibitem[{Segura {et~al.}(2003)Segura, Krelove, Kasting, Sommerlatt, Meadows, Crisp, Cohen, \& Mlawer}]{Seguraetal.2003}
Segura, A., Krelove, K., Kasting, J.~F., {et~al.} 2003, Astrobiology, 3, 689

\bibitem[{Smith {et~al.}(1991)Smith, Yoshino, Parkinson, Ito, \& Stark}]{Smithetal.1991}
Smith, P.~L., Yoshino, K., Parkinson, W.~H., Ito, K., \& Stark, G. 1991, Journal of Geophysical Research: Planets, 96, 17529

\bibitem[{Stern {et~al.}(2015)Stern, Bagenal, Ennico, Gladstone, Grundy, McKinnon, Moore, Olkin, Spencer, Weaver, Young, Andert, Andrews, Banks, Bauer, Bauman, Barnouin, Bedini, Beisser, Beyer, Bhaskaran, Binzel, Birath, Bird, Bogan, Bowman, Bray, Brozovic, Bryan, Buckley, Buie, Buratti, Bushman, Calloway, Carcich, Cheng, Conard, Conrad, Cook, Cruikshank, Custodio, Ore, Deboy, Dischner, Dumont, Earle, Elliott, Ercol, Ernst, Finley, Flanigan, Fountain, Freeze, Greathouse, Green, Guo, Hahn, Hamilton, Hamilton, Hanley, Harch, Hart, Hersman, Hill, Hill, Hinson, Holdridge, Horanyi, Howard, Howett, Jackman, Jacobson, Jennings, Kammer, Kang, Kaufmann, Kollmann, Krimigis, Kusnierkiewicz, Lauer, Lee, Lindstrom, Linscott, Lisse, Lunsford, Mallder, Martin, McComas, McNutt, Mehoke, Mehoke, Melin, Mutchler, Nelson, Nimmo, Nunez, Ocampo, Owen, Paetzold, Page, Parker, Parker, Pelletier, Peterson, Pinkine, Piquette, Porter, Protopapa, Redfern, Reitsema, Reuter, Roberts, Robbins, Rogers, Rose, Runyon, Retherford,
  Ryschkewitsch, Schenk, Schindhelm, Sepan, Showalter, Singer, Soluri, Stanbridge, Steffl, Strobel, Stryk, Summers, Szalay, Tapley, Taylor, Taylor, Throop, Tsang, Tyler, Umurhan, Verbiscer, Versteeg, Vincent, Webbert, Weidner, Weigle, White, Whittenburg, Williams, Williams, Williams, Woods, Zangari, \& Zirnstein}]{Sternetal2015}
Stern, S.~A., Bagenal, F., Ennico, K., {et~al.} 2015, Science, 350, aad1815

\bibitem[{Suto \& Lee(1984)}]{Sutoetal.1984}
Suto, M. \& Lee, L.~C. 1984, The Journal of Chemical Physics, 80, 4824

\bibitem[{Tennyson {et~al.}(2024)Tennyson, Yurchenko, Zhang, Bowesman, Brady, Buldyreva, Chubb, Gamache, Gorman, Guest, Hill, Kefala, Lynas-Gray, Mellor, McKemmish, Mitev, Mizus, Owens, Peng, Perri, Pezzella, Polyansky, Qu, Semenov, Smola, Solokov, Somogyi, Upadhyay, Wright, \& Zobov}]{Tennysonetal2024}
Tennyson, J., Yurchenko, S.~N., Zhang, J., {et~al.} 2024, Journal of Quantitative Spectroscopy and Radiative Transfer, 326, 109083

\bibitem[{Thuillier {et~al.}(2004)Thuillier, Floyd, Woods, Cebula, Hilsenrath, Hersé, \& Labs}]{Thuillieretal.2004}
Thuillier, G., Floyd, L., Woods, T.~N., {et~al.} 2004, Solar Irradiance Reference Spectra (American Geophysical Union (AGU)), 171--194

\bibitem[{Tsai {et~al.}(2023)Tsai, Lee, Powell, Gao, Zhang, Moses, Hébrard, Venot, Parmentier, Jordan, Hu, Alam, Alderson, Batalha, Bean, Benneke, Bierson, Brady, Carone, Carter, Chubb, Inglis, Leconte, Line, López-Morales, Miguel, Molaverdikhani, Rustamkulov, Sing, Stevenson, Wakeford, Yang, Aggarwal, Baeyens, Barat, de~Val-Borro, Daylan, Fortney, France, Goyal, Grant, Kirk, Kreidberg, Louca, Moran, Mukherjee, Nasedkin, Ohno, Rackham, Redfield, Taylor, Tremblin, Visscher, Wallack, Welbanks, Youngblood, Ahrer, Batalha, Behr, Berta-Thompson, Blecic, Casewell, Crossfield, Crouzet, Cubillos, Decin, Désert, Feinstein, Gibson, Harrington, Heng, Henning, Kempton, Krick, Lagage, Lendl, Lothringer, Mansfield, Mayne, Mikal-Evans, Palle, Schlawin, Shorttle, Wheatley, \& Yurchenko}]{Tsaietal.2023}
Tsai, S.-M., Lee, E. K.~H., Powell, D., {et~al.} 2023, Nature, 617, 483

\bibitem[{Tsai {et~al.}(2021)Tsai, Malik, Kitzmann, Lyons, Fateev, Lee, \& Heng}]{Tsaietal.2021}
Tsai, S.-M., Malik, M., Kitzmann, D., {et~al.} 2021, The Astrophysical Journal, 923, 264

\bibitem[{Van~Craen {et~al.}(1985)Van~Craen, Herman, Colin, \& Watson}]{VanCraenetal.1985}
Van~Craen, J.~C., Herman, M., Colin, R., \& Watson, J. K.~G. 1985, Journal of Molecular Spectroscopy, 111, 185

\bibitem[{Van~Craen {et~al.}(1986)Van~Craen, Herman, Colin, \& Watson}]{VanCraenetal.1986}
Van~Craen, J.~C., Herman, M., Colin, R., \& Watson, J. K.~G. 1986, Journal of Molecular Spectroscopy, 119, 137

\bibitem[{Vattulainen {et~al.}(1997)Vattulainen, Wallenius, Stenberg, Hernberg, \& Linna}]{Vattulainenetal.1997}
Vattulainen, J., Wallenius, L., Stenberg, J., Hernberg, R., \& Linna, V. 1997, Applied Spectroscopy, 51, 1311

\bibitem[{Veillet {et~al.}(2024)Veillet, Venot, Sirjean, Bounaceur, Glaude, Al-Refaie, \& Hébrard}]{Veilletetal.2024}
Veillet, R., Venot, O., Sirjean, B., {et~al.} 2024, A\&A, 682, A52

\bibitem[{Venot {et~al.}(2018)Venot, Bénilan, Fray, Gazeau, Lefèvre, Es-sebbar, Hébrard, Schwell, Bahrini, Montmessin, Lefèvre, \& Waldmann}]{venotetal2018}
Venot, O., Bénilan, Y., Fray, N., {et~al.} 2018, A\&A, 609, A34

\bibitem[{Venot {et~al.}(2013)Venot, Fray, Bénilan, Gazeau, Hébrard, Larcher, Schwell, Dobrijevic, \& Selsis}]{Venotetal.2013}
Venot, O., Fray, N., Bénilan, Y., {et~al.} 2013, A\&A, 551, A131

\bibitem[{Venot {et~al.}(2015)Venot, Hébrard, Agúndez, Decin, \& Bounaceur}]{Venotetal.2015}
Venot, O., Hébrard, E., Agúndez, M., Decin, L., \& Bounaceur, R. 2015, A\&A, 577, A33

\bibitem[{Venot {et~al.}(2020)Venot, Parmentier, Blecic, Cubillos, Waldmann, Changeat, Moses, Tremblin, Crouzet, Gao, Powell, Lagage, Dobbs-Dixon, Steinrueck, Kreidberg, Batalha, Bean, Stevenson, Casewell, \& Carone}]{Venotetal.2020}
Venot, O., Parmentier, V., Blecic, J., {et~al.} 2020, The Astrophysical Journal, 890

\bibitem[{Vuitton {et~al.}(2019)Vuitton, Yelle, Klippenstein, Hörst, \& Lavvas}]{Vuittonetal.2019}
Vuitton, V., Yelle, R.~V., Klippenstein, S.~J., Hörst, S.~M., \& Lavvas, P. 2019, Icarus, 324, 120

\bibitem[{Watson {et~al.}(1982)Watson, Herman, Van~Craen, \& Colin}]{Watsonetal.1982}
Watson, J. K.~G., Herman, M., Van~Craen, J.~C., \& Colin, R. 1982, Journal of Molecular Spectroscopy, 95, 101

\bibitem[{Weng {et~al.}(2021)Weng, Li, Aldén, \& Li}]{Wengetal.2021}
Weng, W., Li, S., Aldén, M., \& Li, Z. 2021, Applied Spectroscopy, 75, 1168

\bibitem[{Wu {et~al.}(2001)Wu, Chen, \& Judge}]{Wuetal.2001}
Wu, C. Y.~R., Chen, F.~Z., \& Judge, D.~L. 2001, Journal of Geophysical Research: Planets, 106, 7629

\bibitem[{Wu {et~al.}(2004)Wu, Chen, \& Judge}]{Wuetal.2004}
Wu, C. Y.~R., Chen, F.~Z., \& Judge, D.~L. 2004, Journal of Geophysical Research: Planets, 109

\bibitem[{Wu {et~al.}(1989)Wu, Chien, Liu, Judge, \& Caldwell}]{Wuetal.1989}
Wu, C. Y.~R., Chien, T.~S., Liu, G.~S., Judge, D.~L., \& Caldwell, J.~J. 1989, The Journal of Chemical Physics, 91, 272

\bibitem[{Wu {et~al.}(2000)Wu, Yang, Chen, Judge, Caldwell, \& Trafton}]{Wuetal.2000}
Wu, R. C.~Y., Yang, B.~W., Chen, F.~Z., {et~al.} 2000, Icarus, 145, 289

\bibitem[{Wu {et~al.}(2007)Wu, Lu, Chen, Cheng, Lee, \& Lee}]{Wuetal.2007}
Wu, Y.-J., Lu, H.-C., Chen, H.-K., {et~al.} 2007, The Journal of chemical physics, 127, 154311

\bibitem[{Zabeti {et~al.}(2017)Zabeti, Fikri, \& Schulz}]{Zabetietal.2017}
Zabeti, S., Fikri, M., \& Schulz, C. 2017, Proceedings of the Combustion Institute, 36, 4469

\bibitem[{Zádor {et~al.}(2017)Zádor, Fellows, \& Miller}]{Zadoretal.2017}
Zádor, J., Fellows, M.~D., \& Miller, J.~A. 2017, The Journal of Physical Chemistry A, 121, 4203, doi: 10.1021/acs.jpca.7b03040

\end{thebibliography}

\end{document}